\def\RE {I\kern-6pt R    }
\def\Z  {Z\kern-13pt Z   }
\def\be {\begin{equation}}
\def\ee {\end{equation}  }
\def\beq{\begin{eqnarray}} 
\def\eeq{\end{eqnarray}  }

\def\bi {\begin{itemize} }

\def\ei {\end{itemize}   }
\def\kkbar{{$K{\bar K}$}}
\def\kbar   {$K^{^{^{_{\! \! \! \! \! \! \rule{2.4mm}{0.25mm}}}}}$ }

\newcommand{\scri}{\mathcal {J}}
\documentclass[prd,aps,showpacs,draft,nofootinbib]{revtex4}

\input epsf

\begin{document}

%@twocolumnfalse\endcsname

\title{Fine Structure of Oscillons in the \\ Spherically Symmetric 
$\phi^4$ Klein-Gordon Model}

%\author{Ethan P. Honda and Matthew W. Choptuik}
%\address{Center for Relativity,
%         The University of Texas at Austin,
%         Austin, TX 78712-1081}

\author{Ethan P. Honda}
%\footnote{Actually with the Center for Relativity,
%The University of Texas at Austin during the writing of this paper.}
\affiliation{%Signal Physics Laboratory,
         Applied Research Laboratories,
         The University of Texas at Austin,
         Austin, TX 78758}

\author{Matthew W. Choptuik}
\affiliation{CIAR Cosmology \& Gravity Program \\
         Department of Physics and Astronomy,
         University of British Columbia,
         Vancouver BC, V6T 1Z1\\
         Center for Relativity, University of Texas at Austin, TX 78712-1081 USA}

\begin{abstract}
We present results from a study of the fine structure of 
oscillon dynamics in the 3+1 spherically symmetric Klein-Gordon 
model with a symmetric double-well potential.
We show that in addition to the previously understood longevity
of oscillons, there exists a resonant (and critical) behavior 
which exhibits a time-scaling law.  
The mode structure of the critical solutions is examined, and 
we also show that the upper-bound to oscillon 
formation (in $r_0$ space) is either non-existent or higher than
previously believed.  Our results are generated using a novel 
technique for implementing non-reflecting boundary conditions 
in the finite difference solution of wave equations.  The
method uses a coordinate transformation which blue-shifts and 
``freezes'' outgoing radiation. The frozen radiation is then 
annihilated via dissipation explicitly added to the 
finite-difference scheme, with very little reflection into 
the interior of the computational domain.
\end{abstract}

\maketitle

\vskip2pc

%%%%%%%%%%%%%%%%%%%%%%%%%%%%%%%%%%%%%%%%%%%%%%%%%%%%%%%%%%%%%%%%
\section{Introduction}
\label{sec:tex_introduction}
%%%%%%%%%%%%%%%%%%%%%%%%%%%%%%%%%%%%%%%%%%%%%%%%%%%%%%%%%%%%%%%%

%There is a rich history in the last thirty years in the study 
%of special solutions to non-linear equations 

There is a long history in physics and mathematics of trying to
find new non-trivial solutions to nonlinear wave equations.
The literature on the subject goes back at least as far as 1845 when
J.~Scott Russell
published a paper about a surface wave he witnessed travel for almost two
miles in a shallow water channel
(the first scientifically reported soliton)\cite{Russell}.
Since then there has been much effort directed towards understanding 
{\it stable} localized solutions to nonlinear wave equations: 
the classical kink-soliton, topological defects (monopoles, 
cosmic strings, and domain walls)\cite{Vilenkin}, and nontopological
defects (such as Q-balls)\cite{Friedberg,Coleman}, are but a few 
examples.
However, localized but {\it unstable} solutions have been discussed 
much less frequently and in this paper we focus our attention on one such 
solution, the oscillon.

%Since then there has been much effort directed towards understanding 
%{\it stable} localized solutions to nonlinear wave equations. 
%Some widely known and studied localized and stable solutions 
%include the classical kink-soliton, nontopological solitons (like 
%Q-balls), and topological defects (such as monopoles and vortices). 
%Localized but unstable solutions have been discussed much less 
%frequently, and in this paper we focus our attention on on such 
%solution, the oscillon.
%In this paper we discuss one such solution
%to the nonlinear Klein-Gordon equation.
%Oscillons are seen in many aspects of physics today from vibrating 
%granular material to sub-critical inflationary bubbles.   
%(ie. with $\phi^2$ and $\phi^4$ self-interaction)
%They are typically seen
%resulting from collapsing ``bubbles'' (ie. bubbles that do not have 
%enough energy 
%to expand and contribute to a cosmological phase transition).

The definition of oscillon varies slightly depending 
on context, but here we refer to localized, time-dependent, unstable,
spherically symmetric solutions to the nonlinear 
Klein-Gordon equation.
Although oscillons are unstable, their lifetimes are long compared to 
a dynamical time.
Oscillons were originally discovered by Bogolubsky and Makhan'kov 
\cite{BogMak1,BogMak2} (who called them ``pulsons''), and were later 
studied in more detail by Gleiser \cite{gleiser0} and by 
Copeland {\em et al}~\cite{gleiser1}.

Oscillons can be formed via the collapse of field configurations (initial data) 
that interpolate between two vacuum states 
($\phi_+$ and $\phi_-$) of a symmetric
double well potential (SDWP)\footnote{Asymmetric
double well potentials can also produce oscillons, but here we consider 
only the SDWP.}.
In spherical symmetry, such a configuration 
is a bubble, where the interpolating
region is the bubble ``wall'' that separates the two vacuum states
at some characteristic radius (where in this work we always use 
$\phi_-$ as the large-$r$ vacuum state).  An oscillon formed
this way typically has three distinct stages in its evolution.  
First, immediately following the bubble collapse a large percentage
of its energy is radiated away.  
As will be discussed below,
this can happen 
either through localized field oscillations,
or through bounces reminiscent of the 1+1 dimensional kink-antikink, 
(\kkbar), scattering~\cite{1dkak_campbell}.  
After the initial radiative phase, the solution settles into the 
oscillon stage.  
Here the field is localized with a shape roughly that of an origin-centered
gaussian, with the field value asymptotically approaching the large-$r$
vacuum state. 
%%%%%%%%%%%%%%%%%%%%%%%%%%%%%%%%%%%%%%%%%%%%%%%%%%%%%%%%%%%%%%%%%%%
Due to the asymmetry of the potential about either {\it minimum},
the field oscillates about $\phi_-$ (the large-$r$ vacuum)
such that the time-averaged value of the field lies between 
the two vacua, ie. $\phi_- < \langle\phi\rangle < \phi_+$ where 
$\langle\,\cdots\,\rangle \equiv T^{-1}\int_0^T \, \cdots \, dt$ as 
in \cite{gleiser1}.
%%%%%%%%%%%%%%%%%%%%%%%%%%%%%%%%%%%%%%%%%%%%%%%%%%%%%%%%%%%%%%%%%%%
For typical initial field configurations
the energy of the oscillon is slowly radiated away, approaching a 
specific ``plateau'' value.
In the third and final stage of evolution, the oscillon stops oscillating 
and disperses, radiating away its remaining energy. 

Much of the original excitement about oscillons 
arose from the fact that their long lifetimes
could potentially alter the dynamics of a cosmological phase transition.
However, since oscillons {\it are} unstable, their ability to affect such
a phase transition depends crucially on their lifetimes.
Previous studies by Copeland {\em et al}~\cite{gleiser1}, used dynamical 
grid methods to study oscillon dynamics and treated the
initial radius, $r_0$, of the bubble (shell of radiation), as a free 
parameter.  These studies showed
not only that oscillon lifetimes can be comparable to the
age of the universe (at the GUT scale), but 
that oscillons are formed from a wide range of initial bubble radii.
However, the computational demands of the dynamical grid 
methods employed in~\cite{gleiser1}
prevented a detailed study of the $r_0$ parameter space.

A key problem in the accurate, long-time simulation of oscillons is 
the treatment of boundary conditions at the outer edge, $r = r_{\rm max}$, 
of the computational domain. 
It is standard practice in the computational solution of
nonlinear field equations to use finite difference techniques
applied to functions defined on a lattice of gridpoints.  
If a static, finite-sized domain is used; 
i.e. if $r_{\rm max}$ is fixed,  then one needs to employ a method
which minimizes the amount of radiation (energy) which is artificially
reflected at $r=r_{\rm max}$.  With {\em massless} scalar fields, 
and in spherical symmetry, this can be done quite easily simply by 
imposing a discrete version of an ``outgoing radiation'', or Sommerfeld,
condition.  However, for the case of {\em massive} scalar fields, or 
more generally, for fields with non-trivial dispersion relations, 
the Sommerfeld condition is only approximate, and its use generically 
results in significant reflection at $r=r_{\rm max}$, and subsequent 
contamination of the interior solution.

A surefire fix for the outer-boundary problem is to use 
a dynamically growing grid (as in~\cite{gleiser1}), so 
that $r_{\rm max} = r_{\rm max}(t)$, and 
lattice points are continuously added to extend the computational domain
as needed.
Alternatively, compactified coordinates, or coordinates which propagate 
outwards faster than any characteristic speed in the problem can be 
used, but in these cases, new gridpoints also need to
be continuously added to the mesh in order to maintain adequate resolution
of solution features.  These methods are somewhat more efficient
than the use of a static mesh with $r_{\rm max}$ chosen so that no 
signals reach the outer boundary during the integration period of 
interest, $T$.  However, for long-lived solutions, the mesh soon becomes 
quite large, and the computation time
tends to be proportional to $T^2$.

Recently, 
Gleiser and Sornborger \cite{newgleiser} 
introduced an {\it adiabatic damping method}
%\footnote{Similar to sponge-filters.} 
which adds an explicit damping term to the equations 
of motion, and which has been shown to absorb outgoing massive 
radiation extremely well in 1D (spherical) and 2D (cylindrical)
simulations.  Here we present an alternate 
approach for dealing with outgoing 
massive scalar fields which is quite general
and quite different from previously used methods of which we 
are aware.
The technique involves the use of a specially chosen coordinate system 
that ``freezes'' and blue-shifts 
%%%%%%%%%%%%%%%%%%%%%%%%%%%%%%%%%%%%%%%%%%%%%%%%%%%%%%%%%%%%%%%%%%%%%%%
outgoing
%%%%%%%%%%%%%%%%%%%%%%%%%%%%%%%%%%%%%%%%%%%%%%%%%%%%%%%%%%%%%%%%%%%%%%%
radiation in a relatively thin layer
located away from the central region where the dynamics of principal 
interest unfold.  The addition of a standard type of finite-difference
dissipation~\cite{ko} then ``quenches'' the blue-shifted, frozen radiation, 
and very little energy is reflected back into the interior region.
This approach, like that described in~\cite{newgleiser},
has the advantage that a static and uniform finite-difference mesh can be 
used, so that computational time scales {\em linearly} with the integration
period, $T$.

Our new technique was crucially important to our discovery and detailed 
study of fine structure in a well-known (and still much studied) nonlinear 
system.  Specifically, we have found strong evidence for a family of resonant
oscillon solutions in the SDWP model.  Each of these solutions 
appears to possess
a {\em single} unstable mode in perturbation theory, and by tuning the 
family parameter, $r_0$, in the vicinity of a specific resonance, we 
can tune away that mode, producing oscillons which live longer and longer
as we tune closer and closer to the precise resonant value, $r_0^\star$.  
This leads to a view of oscillons as being analogous to the Type I 
critical solutions which have been discovered in the context of 
gravitational collapse~\cite{TypeI}, 
and as in that case, we find compelling evidence 
for power-law scaling of the oscillon lifetime, $\tau$: 
\begin{equation}
\tau \,\sim\, c_r \vert r_0 - r_0^\star \vert^{\gamma_r}
\end{equation}
where $c_r$ is an overall scale factor set by the particular resonance,
and $\gamma_r$ is a resonance-{\em dependent} exponent which is presumably 
the reciprocal Lyapounov exponent associated with the resonance's single
unstable mode.  

In addition, contrary to previous claims~\cite{gleiser1,newgleiser}, 
we see no hard evidence for an upper bound on $r_0$, beyond which 
oscillons are no longer generated via collapse of gaussian data. 
In particular we find strong evidence for resonances for $r_0 \gtrsim 6.5$, 
well above the limit $r_0 \simeq 4.2$ quoted in~\cite{gleiser1,newgleiser}.
Moreover, we relate the existence of these ``large-$r_0$'' resonances 
to the ``bouncing'' behaviour observed in the 1+1 kink-antikink study 
of Campbell {\em et al}~\cite{1dkak_campbell}.

%%%%%%%%%%%%%%%%%%%%%%%%%%%%%%%%%%%%%%%%%%%%%%%%%%%%%%%%%%%%%%%%

The remainder of the paper is organized as follows:
In Section 2 we introduce a new coordinate system 
in which to solve non-linear wave equations using finite 
differences.  We examine the conformal structure 
induced by our new coordinates, as well as the characteristics
of the resulting wave equation.
In Section 3 we discuss the new properties of oscillons 
which were discovered during our study.
In particular, we observe resonances in the 
parameter space which obey a time-scaling law, and we construct 
a sample resonant solution via a non-radiative {\em ansatz} (Sections 3a and 
3b, respectively).  
Finally, in Section 3c we discuss oscillons and resonant solutions
found outside the bounds of the parameter space previously explored. 
Section 4 summarizes our results and is followed by
two appendices which discuss the details of the
finite difference equations (appendix A) and the testing of the 
code (appendix B).

%%%%%%%%%%%%%%%%%%%%%%%%%%%%%%%%%%%%%%%%%%%%%%%%%%%%%%%%%%%%%%%%
\section{The Klein-Gordon Equation in MIB Coordinates}
\label{sec:tex_eqns}
%%%%%%%%%%%%%%%%%%%%%%%%%%%%%%%%%%%%%%%%%%%%%%%%%%%%%%%%%%%%%%%%

We are interested in the self-interacting scalar field theory 
%with a symmetric double-well potential (SDWP) 
described by the (3+1)-dimensional action
\begin{equation} \label{eq:action}
S[\phi] = \int d^4x\,\sqrt{|g|}\,\left(
-\frac{1}{2} g^{\mu\nu}\nabla_\mu\phi \nabla_\nu\phi - V(\phi)
% \frac{1}{4}\left(\phi^2 -1 \right)^2
\right)
\end{equation}
where we take $V(\phi)$ to be a symmetric double well potential
(SDWP)
\footnote{This is identical to using $\displaystyle{V(\phi) = 
\frac{\lambda}{4}\left(
\phi^2 - \frac{m^2}{\lambda}\right)^2}$ and introducing dimensionless 
variables $r= \tilde{r} m$, $t = \tilde{t} m$, and 
$\chi=\frac{\sqrt{\lambda}}{m}\phi$.},
\begin{equation}
V_S(\phi) = \frac{1}{4}\left(\phi^2 -1 \right)^2
\end{equation}
%or an asymmetric double well potential (ADWP), 
%$\displaystyle{V_A(\phi) = \frac{1}{4}\phi^4 - (1+\epsilon)\phi^3 + \phi^2}$
%\footnote{This becomes a SDWP with minima at $\phi=0$ and $\phi=2$ when 
%$\epsilon=0$},
and
$g_{\mu\nu}$ to be the metric of flat spacetime in spherical symmetry, written 
in standard spherical polar coordinates $(\tilde{t}, \tilde{r}, \tilde{\theta},
\tilde{\varphi})$:
\begin{equation}
d\tilde{s}^2 = -d\tilde{t}^2 + d\tilde{r}^2 + 
                 \tilde{r}^2 \left( d\tilde{\theta}^2 + 
                     \sin^2\tilde{\theta}\,d\tilde{\varphi}^2 \right)
\end{equation}
We now introduce a new radial coordinate, $r$, which 
interpolates between the old radial coordinate, $\tilde{r}$,  at small 
$\tilde{r}$
and an outgoing null coordinate at large $\tilde{r}$.  Specifically, we 
take
\begin{eqnarray}
\tilde{t} &=& t \\ 
\tilde{r} &=& r + f(r) t \\
\tilde{\theta} &=& \theta \\
\tilde{\varphi} &=& \varphi 
\label{eq:transformation}
\end{eqnarray}
where $f(r)$ is a {\it monotonically increasing} function which smoothly 
interpolates
between $\approx 0$ and $\approx 1$ at some characteristic cutoff, 
$r_{\rm c}$, so that $f(r) \to 0$ for $r \ll r_c$, and $f(r) \to 1$ for 
$r \gg r_c$.
%We call these coordinates ($t$,$r$) asymptotically outgoing null (AON) 
%coordinates.  
We call ($t$,$r$) {\em monotonically increasingly boosted} (MIB) 
coordinates.  
The MIB system reduces to the original spherical 
coordinates, ($\tilde{t}$,$\tilde{r}$), for $r\ll r_{\rm c}$, 
but as discussed below, in the $r>r_{\rm c}$ region,
both outgoing {\em and} ingoing (from $r\gg r_c$) radiation tends to 
be ``frozen'' in the transition layer, $r\approx r_c$.
%%%%%%%%%%%%%%%%%%%%%%%%%%%%%%%%%%%%%%%%%%%%%%%%%%%%%%%%%%%%%%%%%%%%%%
Furthermore, since the outgoing radiation is blue-shifted as it propagates 
into the transition region, $r \sim r_c$, application of standard 
finite-difference dissipation operators can then quench it
with minimal reflection back into the interior of the 
computational domain.

In general, MIB coordinates will not cover all of the $(\tilde{t},\tilde{r})$
half-plane.
However, given that $f(r)$ is monotonically increasing, the 
determinant of the Jacobian of the transformation 
%($\det J = 1 + f'(r) t$) 
is non-zero for all $t$ such that $t > -\max|f'(r)|$.
Thus, for this range of $t$, the transformation to and from 
the standard spherical coordinate system is well-defined,
and though a coordinate singularity inevitably forms as $t \to -\infty$
(past timelike infinity),
%(at $\tau=-1/f'(u)$) 
this has no effect on the {\em forward evolution} of initial data given at 
$t=0$.  

We also note that in order that our MIB coordinates be regular at 
$r=0$ (so that there is no conical singularity at the origin), we must also
demand that $f(0)=0$.

Our coordinate choice results in the following spherically
symmetric, 3+1, or ADM~\cite{ADM} form:
%\begin{equation}
%\begin{array}{rcl}
%ds^2 &=& \left(-1 + f^2(u)\right)d\tau^2  + 2 f(u)\left(1 + f'(u)\tau\right) d\tau du \\
%     & &   +\left( 1 + f'(u)\tau\right)^2 du^2 + 
%            \left(u + f(u)\tau \right)^2 d\Omega^2
%\end{array}
%\end{equation}
%or, in a more familiar (3+1) form
\begin{equation}\label{eq:shiftmet} 
ds^2=\left(-\alpha^2+a^2\beta^2\right) dt^2 + 2 a^2\beta dt dr + a^2 dr^2 +
r^2b^2 \left( d\theta^2 + \sin^2\theta\,d\varphi^2\right)
\end{equation}
where
\be
\begin{array}{rclcrcl}
a(t,r)&=& 1+f'(r)t & \ \  & b(t,r)  &=& \displaystyle{1 + f(r)\frac{t}{r}} \\
\vspace{-0.1in} \\
\alpha(t,r)&=& 1 & \ \  & \beta(t,r) &=& \displaystyle{\frac{f(r)}{1 + f'(r)t}}. 
\\
\end{array}
 \label{eq:ADM_auxvars}
\end{equation}
(In the nomenclature of the ADM formalism, $\alpha$ is the 
{\em lapse function}, while $\beta$ is the radial component of the 
{\em shift vector}.)
In the work which follows, we have
adopted the following specific form for $f(r)$: 
\be \label{eq:fdef} f(r)= \left( 1 + \tanh((r-r_{\rm c})/\delta)\right)/2 + \epsilon, \ee
where 
\begin{equation}
\epsilon = -\left( 1 + \tanh(r_{\rm c}/\delta)\right)/2
\end{equation}
is chosen 
to satisfy the regularity condition at $r=0$.

\begin{figure}
\epsfxsize=12.5cm
\centerline{\epsffile{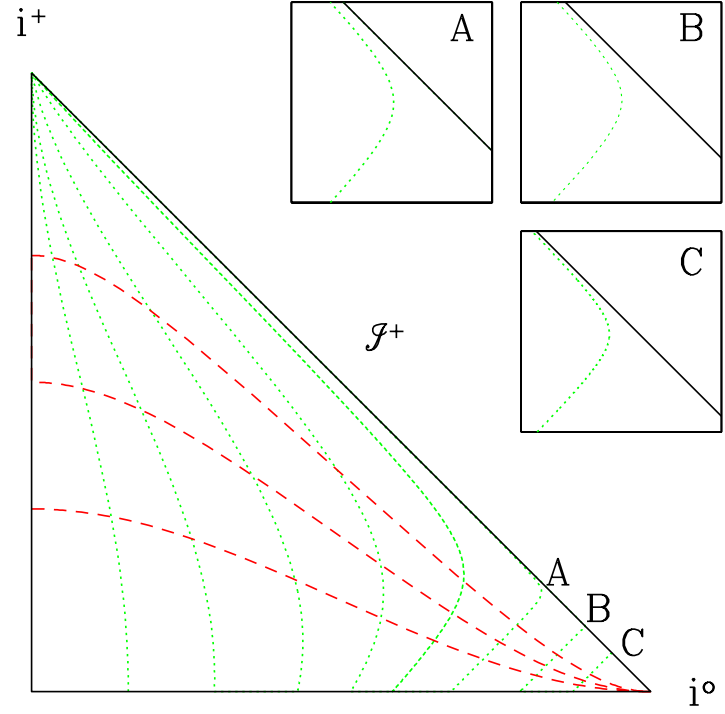}}
\caption[Conformal Diagram for radial MIB coordinates.]
{\small \label{fig:confdiag}
Conformal diagram showing surfaces of constant $r$ (dotted lines)
and lines of constant $t$ (dashed lines).
Lines of constant $t$ look exactly like the constant-$t$
hypersurfaces of Minkowski space, whereas the lines of constant $r$
behave much differently.
For $r>r_c$, it appears as if the constant-$r$ surfaces
are null.  
%This occurs since the coordinates are being shifted outwards
%at {\it nearly} the speed of light.
However, as insets A, B, and C show, the constant-$r$ lines do {\em not}
become null (do not intersect future null infinity), but rather are everywhere
timelike.  
%This is actually easy to understand as the radial
%coordinate is never moved out at the speed of light (not even at
%spatial infinity, because of the $\epsilon$ term in equation
%\ref{eq:fdef}).
}
%\label{fig:confdiag}
\end{figure}

It is now instructive to consider the conformal structure of 
the MIB hypersurfaces.
This is done by applying equations (\ref{eq:transformation}) to 
the standard conformal compactification on Minkowski space,
$\tilde{t} \pm \tilde{r}  = \tan \left( \left( T\pm R\right)/2\right)$
(where $T$ and $R$ are the axes in the conformal diagram,
see \cite{Hawking_Ellis} or \cite{Wald_txt}), 
and then plotting curves of constant $r$ and $t$.
% \label{fig:confdiag} 
\noindent
The constant-$t$ hypersurfaces are 
everywhere spacelike and all reach spatial infinity, ${\rm i^o}$.
Although constant-$r$ surfaces for $r > r_{\rm c}$ appear at 
first glance to be null,
a closer look (see insets of Fig.
\ref{fig:confdiag}) reveals that they are 
indeed everywhere timelike and do not ever reach future null
infinity, {${\scri^+}$}.

The equation of motion for the scalar field which results from the 
action (\ref{eq:action}) is
\begin{equation} \label{eq:boxphi}
\frac{1}{\sqrt{|g|}}\partial_\mu\left( \sqrt{|g|} g^{\mu\nu} 
\partial_\nu\phi \right) = \phi\left(\phi^2 -1 \right)
\end{equation}
which with (\ref{eq:shiftmet}), (\ref{eq:ADM_auxvars}),
$\Pi \equiv a\left( \partial_t{\phi} - \beta \partial_r\phi\right)/\alpha$,
and $\Phi \equiv \partial_r\phi$ give  
\beq 
\dot{\Pi}  &=& \frac{1}{r^2b^2}\left( r^2 b^2 \left( \frac{\alpha}{a} \Phi 
+ \beta \Pi \right)\right)'  
- 2 \frac{\dot{b}}{b}\Pi - \alpha a \phi \left( \phi^2 - 1\right)
\label{eq:EOM_Pi}\\
\dot{\Phi} &=& \left( \frac{\alpha}{a} \Pi + \beta \Phi \right)' 
\label{eq:EOM_Phi}\\
\dot{\phi} &=& \frac{\alpha}{a} \Pi + \beta \Phi 
\label{eq:EOM_phi}
\eeq
where \ $\dot{} \equiv \partial_t$ and \ ${}' \equiv\partial_r$. 
These equations are familiar from the ADM formalism as applied 
to the spherically symmetric Klein-Gordon field coupled
to the general relativistic gravitational field~\cite{choptuik:1986}.
However, in the current case, instead of a dynamically evolving 
metric functions, the metric components $a$, $b$, $\alpha$, and $\beta$
are {\em a priori} 
fixed functions of $(t,r)$
that resulted from a coordinate transformation of {\it flat spacetime}.

%%%%%%%%%%%%%%%%%%%%%%%%%%%%%%%%%%%%%%%%%%%%%%%%%%%%%%%%%%%%%%%%

%%%%%%%%%%%%%%%%%%%%%%%%%%%%%%%%%%%%%%%%%%%%%%%%%%%%%%%%%%%%%%%%%%%%%%%%%
Clearly, characteristic speeds for the massless Klein-Gordon 
field ($V(\phi) = 0$)
bound the inward or outward speed (group velocities) of {\em any} radiation 
in a {\em self-interacting} field ($V(\phi) \ne 0$).
Characteristic analysis of the {\em massless} %($V(\phi) = 0$)
Klein-Gordon equation
with metric (\ref{eq:shiftmet}) yields local propagation speeds
\begin{equation}
\lambda_\pm = -\beta \pm \frac{\alpha}{a},
\label{eq:1dchars}
\end{equation}
where $\lambda_+$ and $\lambda_-$ are the outgoing and ingoing characteristic 
speeds, respectively \cite{choptuik:1986,courant:1962}.  
%Clearly, these characteristic 
%speeds bound the inward or outward speed (group velocities) of any radiation 
%in a {\em self-interacting} ($V(\phi) \ne 0$) field.
%%%%%%%%%%%%%%%%%%%%%%%%%%%%%%%%%%%%%%%%%%%%%%%%%%%%%%%%%%%%%%%%%%%%%%%%%
%%%%%%%%%%%%%%%%%%%%%%%%%%%%%%%%%%%%%%%%%%%%%%%%%%%%%%%%%%%%%%%%%%%%%%%%%
For $r\ll r_{\rm c}$, the propagation of scalar radiation in  
$(t,r)$ or $(\tilde{t},\tilde{r})$
coordinates is essentially identical.  However, as illustrated in 
Fig.~\ref{fig:1Dchars_coin}, and as can be deduced from 
equations (\ref{eq:1dchars}) and (\ref{eq:ADM_auxvars}), for
$r\approx r_{\rm c}$
{\em both} the ingoing and the outgoing characteristic velocities go to 
zero as $t\rightarrow \infty$ (as the inverse power of $t$).
%%%%%%%%%%%%%%%%%%%%%%%%%%%%%%%%%%%%%%%%%%%%%%%%%%%%%%%%%%%%%%%%%%%%%%%
%%%%%%%%%%%%%%%%%%%%%%%%%%%%%%%%%%%%%%%%%%%%%%%%%%%%%%%%%%%%%%%%%%%%%%%
Thus, {\em any} radiation incident on this region will effectively be 
trapped, or ``frozen in''.
It is this property
of the MIB system which enables the effective implementation of 
non-reflecting boundary conditions.
As discussed further in Appendix~A, 
an additional key ingredient is the application of Kreiss-Oliger-style 
dissipation~\cite{ko} to the difference equations. This 
dissipation efficiently 
quenches the trapped outgoing radiation, which as mentioned above tends 
to be blue-shifted to the lattice scale on a dynamical time-scale.

Finally, we note that, as is evident from Fig.~\ref{fig:1Dchars_coin},
the ``absorbing layer'' in the MIB system (i.e. the region in which 
the characteristic speeds are $\approx 0$), expands both outward and 
inward as $t$ increases.  This means that for fixed $r_c$, the absorbing 
layer will eventually encroach on the interior region $r \ll r_c$ and 
ruin the calculation.  However, the rate at which the layer expands 
is roughly logarithmic in $t$, so, in practice, this fact should not 
significantly impact the viability of the method. 
For arbitrarily large final integration times, $T$,
computational cost will scale as $T \ln T$. 
However, the calculations described here all used the same values of 
$r_c$ and $r_{\rm max}$, so that for all practical purposes, the 
computational cost is {\em linear} in the integration time.

%\label{fig:1Dchars}
%\label{fig:1Dchars_coin}

%Looking back to equations (\ref{eq:ADM_auxvars}), 
%we see that as $r$ transitions through $r_{\rm c}$, $\beta$ goes 
%from zero to almost one.  It is this ``shift'' term (arising from the new 
%radial coordinate moving out at almost the speed of light) that is 
%responsible for freezing out the outgoing radiation.

%%%%%%%%%%%%%%%%%%%%%%%%%%%%%%%%%%%%%%%%%%%%%%%%%%%%%%%%%%%%%%%%
\section{The Resonant Structure of Oscillons}
\label{sec:tex_fine}
%%%%%%%%%%%%%%%%%%%%%%%%%%%%%%%%%%%%%%%%%%%%%%%%%%%%%%%%%%%%%%%%

Copeland, {\em et al}, showed quite clearly that oscillons formed 
for a wide range of initial bubble radii, $r_0$.
They even caught a glimpse of the fine structure in the model---which 
in large part motivated this study---but they did not explore 
this fine structure of the parameter space in detail.
With the efficiency of our new code, we have been able to explore
parameter space much more thoroughly, which in turn has yielded 
additional insights into the dynamical nature of oscillons.

Following \cite{gleiser0} we use a gaussian profile for initial 
data where the field at the core and outer boundary values are 
set to the vacuum values, $\phi(t,0) \equiv \phi_c=1$ and 
$\phi(t,\infty) \equiv \phi_o=-1$ respectively,
and the field interpolates between them at a characteristic radius,
$r_0$:
\be
\phi(0,r) = 
\phi_o + \left(\phi_c - \phi_o\right) \exp\left( 
- r^2 / r_0^2 \right).
\ee
Keeping 
$\phi_c$ and $\phi_o$ constant, but varying $r_0$, we have a one
parameter family of solutions to explore.
Figure \ref{fig:lifetime1} shows the behavior of oscillon lifetime 
as a  function of $r_0$ in the range $2.0 \le r_0 \le 5.0$.
%\label{fig:lifetime1}
We discuss three main findings that are distinct from 
previous work: the existence of resonances and their 
time scaling properties, 
the mode structure of the resonant solutions, and the 
existence of oscillons outside the  parameter-space region $2\leq r_0\leq 5$.

\subsection{Resonances \& Time Scaling}

In contrast to Fig.~7 of Copeland {\em et al}~\cite{gleiser1},
the most obvious new feature seen in Fig. \ref{fig:lifetime1} 
is the appearance of the 125 resonances which rise above 
the overall lifetime profile.  
These resonances (also seen in Fig.
\ref{fig:life_1449.r0=2.27_2.29})
become visible only after carefully resolving the parameter space.
%\begin{figure}
%\epsfxsize=7.5cm
%\centerline{\epsffile{life_1449.r0=2.27_2.29.eps}}
%\caption[lifetime vs radius, all]
%{
%Plot of Oscillon lifetime versus Initial Bubble Radius
%for  $2.27<r_0<2.29$.
%}
%\label{fig:life_1449.r0=2.27_2.29}
%\end{figure}
%\label{fig:life_1449.r0=2.27_2.29}
Upon fine-tuning $r_0$ to about 1 part in $\sim 10^{14}$ we 
noticed interesting bifurcate behavior about the resonances
(Figure \ref{fig:phenv_energy}, top).
%%%%%%%%%%%%%%%%%%%%%%%%%%%%%%%%%%%%%%%%%%%%%%%%%%%%%%%%%%%%%%%%
%%%%%%%%%%%%%%%%%%%%%%%%%%%%%%%%%%%%%%%%%%%%%%%%%%%%%%%%%%%%%%%%
The field oscillates with a period $T \approx 4.6$ (for {\it all} oscillons)
so the individual oscillations 
cannot be seen in the plot, but it is the lower-frequency modulation 
that is 
of interest here\footnote{In dimensionful coordinates, $\tilde{r}$ and 
$\tilde{t}$, 
the period would be
$\tilde{T} = 4.6m^{-1}$.
In general, to recover proper dimensions, lengths and times are multiplied by 
$m^{-1}$ and energies by $\lambda m^{-1}$.}. 
%the scalar field at the core of the oscillon,
The top figure shows the envelope of $\phi(t,0)$ 
on both sides of a resonance (dotted and solid curves).  We see that the 
large period 
modulation that exists for all typical oscillons disappears late in the
lifetime of the oscillon as $r_0$ is brought closer to a resonant
value, $r^\star_0$.
%%%%%%%%%%%%%%%%%%%%%%%%%%%%%%%%%%%%%%%%%%%%%%%%%%%%%%%%%%%%%%%%
On one side of $r^\star_0$ the modulation returns before the oscillon 
disperses (refered to as {\em supercritical} and shown with the solid curve), 
while on the other side of $r^\star_0$ the modulation does not return 
and the the oscillon simply disperses (refered to as {\em subcritical} 
and shown with the dotted curve).  
For resonances where $r^\star_0 \lesssim 2.84$, 
the subcritcal solutions
appear on the $r_0<r^\star_0$ side of the resonance and the 
supercritical solutions appear on the 
$r_0>r^\star_0$ side of the resonance.
The opposite is true for resonances where $r^\star_0 \gtrsim 2.84$,
ie. the subcritcal solutions
appear on the $r_0>r^\star_0$ side of the resonance and the 
supercritical solutions appear on the 
$r_0<r^\star_0$ side of the resonance. 
This bifurcate behavior does not manifest itself until $r_0$ is quite close
to $r^\star_0$.  
%%%%%%%%%%%%%%%%%%%%%%%%%%%%%%%%%%%%%%%%%%%%%%%%%%%%%%%%%%%%%%%%%
%%%%%%%%%%%%%%%%%%%%%%%%%%%%%%%%%%%%%%%%%%%%%%%%%%%%%%%%%%%%%%%%
In practice then, to locate a resonant value, $r^\star_0$, 
we first maximize the oscillon lifetime using a three point extremization
routine ({\it golden section search} with bracketing interval of 
$\sim\!0.62$, \cite{NumericalRecipes}) until we have computed an 
interval whose end-points exhibit the two distinct behaviours just
described.  Once a resonance has been thus bracketed, we switch to a 
standard bisection search, and subsequently locate the resonance to 
close to machine precision.
%\label{fig:phenv_energy}
%\begin{figure}
%\epsfxsize=7.5cm
%\epsfysize=7.5cm
%\centerline{\epsffile{phenv_energy.eps}}
%\caption[Phi envelope at criticality, radiated power]
%{Envelope of $\phi(t,r\!\!=\!\!0)$ for $r_0\!\pm\!\Delta r_0$
%displaying bifurcate behavior around 
%$r_0 \approx 2.335$ resonance (top, $\Delta r_0 \sim 10^{-14}$). 
%Bottom plot shows the energy radiated out of the gaussian surface
%containing the oscillon. }
%\label{fig:phenv_energy}
%\end{figure}
Although we can see from Fig. \ref{fig:phenv_energy} that
the modulation is directly linked to the resonant solution, 
it is not obvious why this is so.
However, if we look at the relationship between the modulation in the 
field (top) to the power radiated by the oscillon (bottom), 
we see that they are clearly synchronized.  

The behaviour of these resonant solutions may not be surprising to
those familiar with the 1+1 \kkbar\ scattering studied using the
same model~\cite{1dkak_campbell}.  Campbell {\em et al}
showed that 
after the ``prompt radiation'' phase---the initial release
of radiation upon collision of a kink and anti-kink---the remaining radiation 
was emitted from the decay of what they 
referred to as ``shape'' oscillations.  The ``shape modes'' 
were driven by the contribution to the field ``on top'' of the 
$K$ and \kbar soliton solutions.  Since the exact closed-form solution for 
the ideal non-radiative \kkbar\ interaction is not known, initial data
aimed at generating such an interactionm is generally only approximate,
and the ``surplus'' (or deficit) field is responsible for 
exciting the shape modes.  The energy stored in the 
shape modes slowly decays away as the kink and antikink interact
and eventually the solution disperses.

%%%%%%%%%%%%%%%%%%%%%%%%%%%%%%%%%%%%%%%%%%%%%%%%%%%%%%%%%%%%%%%%
%%%%%%%%%%%%%%%%%%%%%%%%%%%%%%%%%%%%%%%%%%%%%%%%%%%%%%%%%%%%%%%%
In our case, we believe the large period modulation represents
the excitation of a similar ``shape mode'' superimposed on
a periodic, non-radiative, localized oscillating
solution. On either side
of a resonance in the $r_0$ parameter space, the solution is on the 
threshold of having one more shape mode oscillation.
If this is the case, then, as we tune $r_0 \to r_0^\star$, we are, in 
effect, tuning away the {\em single} unstable shape mode, and thus 
should expect that the oscillon lifetime obey a scaling law such as that 
seen in Type I solutions in critical gravitational collapse~\cite{TypeI}.
%\label{fig:tscaling}
%\begin{figure}
%\epsfxsize=7.5cm
%\centerline{\epsffile{sample_scaling.eps}}
%\caption{Sample plot of time scaling, $T$ versus $\ln|r_0 - r_0^\star|$ about a resonance.
%The top line displays the scaling behavior for $r_0>r_0^\star$ 
%while the bottom line for $r_0<r_0^\star$.}
%\label{fig:tscaling}
%\end{figure}
Figure \ref{fig:tscaling} shows a plot of oscillon lifetime versus 
\hbox{$\ln|r_0-r_0^\star|$} (for the $r_0=2.335\cdots$ resonance), and we can 
see quite 
clearly that there {\em is} a scaling law, $T\sim\gamma\ln|r_0-r_0^\star|$, 
for the lifetime of the solution as measured on either side of the resonance.
We denote $\gamma_+$ for the scaling exponent on the 
$r_0>r_0^\star$ side, and $\gamma_-$ for the scaling exponent on the 
$r_0<r_0^\star$ side.  
Although we observe lifetime scaling for each resonance, the 
scaling exponent {\em per se} varies from
resonance to resonance;
a plot of the scaling exponents, $\gamma_+$ and $\gamma_-$, 
versus the critical initial
bubble radius can be seen in Fig.~\ref{fig:exponents}.
%%%%%%%%%%%%%%%%%%%%%%%%%%%%%%%%%%%%%%%%%%%%%%%%%%%%%%%%%%%%%%%%
%%%%%%%%%%%%%%%%%%%%%%%%%%%%%%%%%%%%%%%%%%%%%%%%%%%%%%%%%%%%%%%%
For all resonances we find
$\gamma_+\approx\gamma_-$.
%\label{fig:exponents}

Finally we note that, by analogy with the case of Type I critical 
gravitational collapse, we expect that the scaling exponents, $\gamma$, 
are simply the reciprocal Lyapounov exponents associated with each 
resonance's single unstable mode.  In addition we note that,  for any 
resonance, if we were able to {\em infinitely} fine-tune $r_0$ to $r_0^\star$,
we would expect the oscillon lifetime to go to infinity.

%\begin{figure}
%\epsfxsize=7.5cm
%%\centerline{\epsffile{gamma_life_vs_r.eps}}
%\centerline{\epsffile{gamma_vs_r.both.err.eps}}
%\caption{
%Plot of critical exponents for each resonance. 
%The uncertainties are estimated from running the entire parameter
%space surveys at two resolutions.
%}
%\label{fig:exponents}
%\end{figure}

\subsection{Mode Structure}

%strictly 
Assuming that periodic, non-radiative solutions to 
equation~(\ref{eq:boxphi}) exist, we should be able to construct 
them by inserting an 
{\em ansatz} of the form
\be
\phi(t,r) = \phi_0(r) + \sum_{n=1}^{\infty}\phi_n(r) \cos\left(n\omega t\right)
\label{eqn:ansatz}
\ee
in the equations of motion 
and solving the resulting system of ordinary differential equations
obtained from matching $\cos\left(n\omega t\right)$ terms:
\begin{equation}
\begin{array}{rcl}
\left( r^2 \phi_0' \right)'/r^2
&=& \phi_0\left( \phi_0 - 1\right)\left(\phi_0-2\right) \\
&&+ \frac{3}{2} \left( \phi_0 -1\right) \sum\limits_m \left( \phi_m\right)^2\\
&&+ \frac{1}{4}\sum\limits_{m,p,q}\phi_n\phi_p\phi_q\left(
\delta_{m, \pm p \pm q} \right) ,
\end{array}
\label{modestruct0}
\end{equation}
\begin{equation}
\begin{array}{rcl}
\left( r^2 \phi_n' \right)' /r^2
&=& \left( 3 \left(\phi_0-1\right)^2 -
\left( n^2 \omega^2 +1\right)\right)\phi_n\\
&&+\frac{3}{2} \left(\phi_0-1\right) \sum\limits_{p,q}\phi_p\phi_q 
\left(\delta_{n,\pm p \pm q}\right) \\
&&+\frac{1}{4}\sum\limits_{m,p,q}\phi_m\phi_n\phi_q\left(
\delta_{n,\pm m \pm p \pm q}\right) .
\end{array}
\label{modestructn}
\end{equation}
Equations (\ref{modestruct0}) and (\ref{modestructn}) can also be 
obtained by inserting {\em ansatz} (\ref{eqn:ansatz}) into the action 
and varying with respect to the $\phi_n$
\cite{morrisonchat}.
This set of ODEs can be solved by ``shooting'', where the 
quantities $\phi_n(0)$ are the shooting parameters. 
Unfortunately, we were unable to construct a method that 
self-consistently computed $\omega$; the best we could achieve was to 
solve equations (\ref{modestruct0}) and (\ref{modestructn})
for a given $\omega$ which we measured from the PDE solution.  

For ease of comparison of the results obtained from the periodic 
{\em ansatz} with those generated via solution of the PDEs, we 
Fourier decomposed the PDE results.  This was done by taking the solution
%$\phi(t_n,r_i)$ 
during the interval of time when the large period 
modulation disappears ($1200<t<1800$ 
for the oscillon in Fig.~\ref{fig:phenv_energy}, for example)
and constructing FFTs of $\phi$ at each gridpoint, $r_i$.  
%Specifically, at each $r_i$, the amplitude of each Fourier 
%mode was obtained 
%from a FFT which used a time series of length 4096.
Specifically, at each $r_i$, the amplitude of each Fourier 
mode was obtained from a FFT which used a time series, 
%%%%%%%%%%%%%%%%%%%%%%%%%%%%%%%%%%%%%%%%%%%%%%%%%%%%%%%%%%%%%%%%
%%%%%%%%%%%%%%%%%%%%%%%%%%%%%%%%%%%%%%%%%%%%%%%%%%%%%%%%%%%%%%%%
$\phi(t^n,r_i)$, $n = 1, 2, \, \cdots \, 4096$ with 
$t^{n+1}-t^n \equiv \Delta t = {\rm constant}$.
%%%%%%%%%%%%%%%%%%%%%%%%%%%%%%%%%%%%%%%%%%%%%%%%%%%%%%%%%%%%%%%%
%%%%%%%%%%%%%%%%%%%%%%%%%%%%%%%%%%%%%%%%%%%%%%%%%%%%%%%%%%%%%%%%%
%\label{fig:modstruct}
%\begin{figure}
%\epsfxsize=7.5cm
%\centerline{\epsffile{modestruct.eps}}
%\caption{
%Sample $\phi_n(r)$ (for $n=0,1,2,3,4$) obtained 
%from the Fourier decomposed dynamic data (x's)
%overlayed with $\phi_n(r)$ obtained by 
%shooting equations \ref{modestruct0} and \ref{modestructn} 
%(solid curves).
%}
%\label{fig:modstruct}
%\end{figure}
Keeping only the first five modes in the expansion~(\ref{eqn:ansatz}), 
we compare the Fourier decomposed PDE data with the
shooting solution (see Fig. \ref{fig:modstruct}).
It should be noted that 
although the value for $\omega$ was determined from the PDE solution, 
the shooting algorithm still involved a five-dimensional search
for the the shooting parameters, $\phi_n(0)$, $n=0,\cdots,4$.
The close correspondence of the curves shown in Fig~\ref{fig:modstruct}
strongly suggests that the resonant solutions
(ie. in the limit as $r_0\rightarrow r_0^\star$) 
observed in the PDE calculations
are indeed consistent with 
the periodic, non-radiative oscillon
{\em ansatz} (\ref{eqn:ansatz}).

%\label{fig:modrad}
%\begin{figure}
%\epsfxsize=7.5cm
%\centerline{\epsffile{resmod.eps}}
%\caption[Power Spectra]
%{
%Power spectra of the core amplitude, $\phi(t,r\!\!=\!\!0)$, 
%for the oscillons barely above and below the 
%$r_0\approx2.335$ resonance.  
%}
%\label{fig:modrad}
%\end{figure} 

By examining the three most dominant components of the 
power spectrum of $\phi(t,0)$, Fig.~\ref{fig:modrad},
we can see that during the ``no-modulation'' epoch, the 
amplitude of each Fourier mode becomes constant.  Although the 
specific plot is for the core amplitude, $r=0$, we note that this 
behavior holds for all $r$.   Again, this is consistent with the 
view that as we tune $r_0$ to $r_0^\star$, the oscillon phase of 
the evolution is better and better described by a one-mode unstable,
``intermediate attractor''.  As discussed previously,
this is precisely reminiscent of the Type I critical phenomena 
studied in critical gravitational collapse, particularly 
the collapse of a real, massive scalar field as studied by 
Brady {\em et al}~\cite{TypeI}, where the intermediate attractors are 
unstable, periodic, ``oscillon stars'' discovered earlier by 
Seidel and Suen~\cite{SeidelSuen}.

\subsection{(Bounce) Windows to more Oscillons}

Lastly, we consider the existence of oscillons generated by gaussian
initial data with $r_0 \gtrsim 5$.
The oscillons explored by Copeland {\em et al}, were restricted to 
the parameter-space region, $2 \lesssim r_0 \lesssim5$, and in fact 
it was concluded that there was an upper bound, $r_0 \sim 4.2$, 
beyond which evolution of gaussian data would not result in an 
oscillon phase~\cite{newgleiser}. However, we have found that 
oscillons {\em can} form for $r_0 \gtrsim 5$, and that they do 
so by a rather interesting mechanism.  

Again, from the 1+1 dimensional \kkbar\ scattering studies of
Campbell {\em et al}, it is well known that a kink and antikink 
in interaction
often ``bounce'' many times before either dispersing or falling 
into an (unstable) bound state.
Here, a bounce occurs when the kink and antikink reflect off one another, 
stop after propagating a short distance, then recollapse.  

We find that such behaviour occurs in the (3+1) dimensional case as well, 
but now the unstable bound state is an oscillon.
For larger $r_0$, instead of remaining within 
$r\lesssim 2.5$ 
after reflection through $r=0$
(as occurs for $2 \lesssim r_0 \lesssim 5$), 
the bubble wall travels out to 
larger $r$ (typically
$3\lesssim r \lesssim 6$), stops, then recollapses, shedding away large 
amounts of
energy in the process (see Fig. \ref{fig:corebounce}).  %EPH2: added ref
%\label{fig:corebounce} %EPH2: had no label?
%\begin{figure}
%\epsfxsize=7.5cm
%\centerline{\epsffile{chaoscore.by2.eps}}
%\caption[Chaotic behavior of phicore]
%{
%Plot of $\phi(t,r=0)$ 
%for $r_0=7.25$.
%}
%\end{figure}
Thus in this system, as with the 1+1 $K{\bar K}$ model, there are 
regions of parameter space which constitute ``bounce windows''.  Within
such regions, the bounces allow the bubble to radiate away large amounts 
of energy.
The bubble then recollapses, effectively producing a new initial 
configuration (albeit with a different shape) with a smaller effective 
$r_0$.
%In Fig. \ref{fig:lifefine} we see a lifetime 
%profile and its resonances for a typical bounce window.
Within these ``windows'' both oscillons and resonances 
(similar to those
observed for $2\lesssim r_0\lesssim 4.6$)
can be observed
(inset of Fig. \ref{fig:lifefine}).
%\label{fig:lifefine}
%\begin{figure}
%\epsfxsize=7.5cm
%\centerline{\epsffile{life_chaos.eps}}
%\caption[lifetime in Chaotic region]
%{
%Plot of Oscillon lifetime versus Initial Radius of Bubble
%for $4.22<r_0<9$.
%}
%\label{fig:lifefine}
%\end{figure}

%%%%%%%%%%%%%%%%%%%%%%%%%%%%%%%%%%%%%%%%%%%%%%%%%%%%%%%%%%%%%%%%
\section{Conclusions}
\label{sec:tex_conclusions}
%%%%%%%%%%%%%%%%%%%%%%%%%%%%%%%%%%%%%%%%%%%%%%%%%%%%%%%%%%%%%%%%

Using a new technique for implementing non-reflecting boundary 
conditions for finite-differenced evolutions of non-linear 
wave equations, we have conducted an extensive parameter space
survey of bubble dynamics described by a spherically-symmetric 
Klein-Gordon field with a symmetric double-well potential.
We have found that the parameter space of the model exhibits 
resonances, wherein the lifetimes of the intermediate-phase 
``oscillons'' diverge as one approaches a resonance.  We have conjectured 
that these resonances are single-mode unstable solutions, analogous to 
Type I solutions in critical gravitational collapse, and have presented 
evidence that their lifetimes satisfy the type of scaling law which is to be 
expected if this is so.  

In addition, we have independently computed 
resonant solutions starting from an {\em ansatz} of periodicity, 
and have demonstrated good agreement between the solutions thereby
computed, and those generated via finite-difference solution of the 
PDEs.  Finally, we have showed that
oscillons can form from bubbles with energies higher than had 
previously been assumed, through a mechanism analogous to the
bounce windows found in the $1+1$ case of kink-antikink scattering.

We note that the use of MIB or related coordinates, in conjunction 
with finite-difference dissipation techniques, should result in 
a generally-applicable strategy for formulating non-reflecting 
boundary conditions for finite-difference solution of 
wave equations.  The method has already been used in the 
study of axisymmetric oscillon collisions~\cite{Thesis}, and 
attempts are underway to use similar techniques in the context 
of 3-D numerical relativity 
and 2-D and 3-D ocean acoustics.  %EPH: This is true, :), 
%     but we don't have to include it...
%     ...coming to a JASA article near you.

%%%%%%%%%%%%%%%%%%%%%%%%%%%%%%%%%%%%%%%%%%%%%%%%%%%%%%%%%%%%%%%%
\section{Acknowledgments}
\label{sec:ack}
%%%%%%%%%%%%%%%%%%%%%%%%%%%%%%%%%%%%%%%%%%%%%%%%%%%%%%%%%%%%%%%%

We thank Philip J. Morrison for useful discussions on the 
exact non-radiative oscillon solutions.  We also thank Marcelo 
Gleiser and Andrew Sornborger for useful input on many 
oscillon issues. 
This research was supported by NSF grant PHY9722088, 
a Texas Advanced Research Projects Grant, and by NSERC.
The bulk of the computations described here were carried out
on the {\tt vn.physics.ubc.ca} Beowulf cluster, which was
funded by the Canadian Foundation for Innovation, with
operations support from NSERC, and from the Canadian Institute for
Advanced Research.
Some computations were also carried out using the
Texas Advanced Computing Center's
SGI Cray SV-1 {\tt aurora.hpc.utexas.edu} and SGI Cray T3E
{\tt lonestar.hpc.utexas.edu}.

\newpage
%%%%%%%%%%%%%%%%%%%%%%%%%%%%%%%%%%%%%%%%%%%%%%%%%%%%%%%%%%%%%%%%
\appendix
\section{Finite Difference Equations}
\label{app:FDE}
%%%%%%%%%%%%%%%%%%%%%%%%%%%%%%%%%%%%%%%%%%%%%%%%%%%%%%%%%%%%%%%%
%%%%%%%%%%%%%%%%%%%%%%%%%%%%%%%%%%%%%%%%%%%%%%%%%%%%%%%%%%%%%%%%

Equations (\ref{eq:EOM_Pi},\ref{eq:EOM_Phi},\ref{eq:EOM_phi}) 
are solved using two-level second order (in both space and time) 
finite difference approximations on a static uniform spatial mesh:
\begin{equation}
r_i = \left( i - 1 \right)\,\Delta r, \quad i = 1, 2, \cdots I
\end{equation}
where $I$ is the total number of mesh points 
\begin{equation}
\Delta r = \frac{r_{\rm max}}{I - 1} \, .
\end{equation}
The scale of discretization is set by $\Delta r$ and $\Delta t =
\lambda \Delta r$, where we fixed the Courant factor, $\lambda$,
to $0.5$ as we changed the base discretization.

%\label{tab:fdop}
Using the operators from Table \ref{tab:fdop}, 
$\partial_r\tilde{r}=a$, 
\hbox{$\partial_r=nr^{n-1}\partial_{r^n}$}, 
and $rb=\tilde{r}$,
the difference equations applied in the interior of the mesh, 
$i = 2, 3, \cdots I - 1$, are
\begin{equation}
\begin{array}{rcl}
\displaystyle{ \Delta^d_t \Pi^n_i } &=& 
\displaystyle{ 3\mu_t\left[a \Delta_{\tilde{r}^3}\left(
\tilde{r}^2 \left( \frac{\alpha}{a}\Phi + \beta\Pi\right)\right)
\right]^n_i} \\
&& 
\displaystyle{-2\mu_t\left(\frac{\dot{b}}{b} \Pi
-\alpha a \phi\left(\phi^2 - 1\right)\right)^n_i},
\end{array}
\end{equation}
\begin{equation}
\Delta^d_t\Phi^n_i = \mu_t\Delta_r\left( 
\frac{\alpha}{a}\Pi + \beta\Phi
\right)^n_i,
\end{equation}
\begin{equation}
\Delta^d_t\phi^n_i = \mu_t\left( 
\frac{\alpha}{a}\Pi + \beta\Phi
\right)^n_i.
\end{equation}
These equations are solved using an iterative scheme and explicit 
dissipation of the type advocated by Kreiss and Oliger~\cite{ko}.
The dissipative term, incorporated in the operator $\Delta^d_t$,
is essentially a fourth spatial 
derivative multiplied by $\left(\Delta r\right)^3$ so that the 
truncation error of the difference scheme remains 
$O(\Delta r^2,\Delta t^2)$.
The temporal difference operator, $\Delta^d_t$, is used as an approximation
to $\partial_t$ everywhere in the interior of the computational domain,
except for next-to-extremal points, where  
$\Delta_t$ is used because the grid values $f^n_{i+2}$ or 
$f^n_{i-2}$ are not defined.  

At the inner boundary, $r=0$, we use $O(\Delta r^2)$ forward 
spatial differences to evolve $\Pi$
\begin{equation}
\mu_t\left( \Delta^f_i\Pi  -
\displaystyle{ \frac{a'}{a} \Pi }\right)^n_1 =0.
\end{equation}
whereas $\Phi^n_1$  is fixed by regularity:
\begin{equation}
\Phi^n_1 = 0.
\end{equation}

To update $\phi$, we use a discrete versions of the equation 
for ${\dot \phi}$  which follows from the definition 
of $\Pi$:
\begin{equation}
\Delta_t\phi = \mu_t\left( 
\frac{\alpha \Pi}{a} + \beta\Phi
\right)^n_i \quad\quad i = 1, 2, \cdots I
\end{equation}
At the outer boundary, $r=r_{\rm max}$, our specific choice of boundary 
conditions and discretizations thereof have little
impact; due to the use of MIB coordinates and Kreiss-Oliger 
dissipation, almost none of the outgoing scalar field reaches 
the outer edge of the computational domain.
Nevertheless, we imposed discrete versions of the usual Sommerfeld conditions 
for a {\em massless} scalar field
on $\Pi$ and $\Phi$:
\begin{equation}
\Delta_t \Pi^n_i + \mu_t \left( 
\Delta^b_r\Pi + \frac{\Pi}{r}
\right)^n_I=0,
\end{equation}
\begin{equation}
\Delta_t \Phi^n_i + \mu_t \left( 
\Delta^b_r\Phi + \frac{\Phi}{r}
\right)^n_I=0,
\end{equation}

%%%%%%%%%%%%%%%%%%%%%%%%%%%%%%%%%%%%%%%%%%%%%%%%%%%%%%%%%%%%%%%%
\section{Testing the MIB Code}
\label{app:codetest}
%%%%%%%%%%%%%%%%%%%%%%%%%%%%%%%%%%%%%%%%%%%%%%%%%%%%%%%%%%%%%%%%

One might think that ``freezing'' outgoing radiation 
on a static uniform mesh  
would lead to a ``bunching-up'' of the wave-train from 
the oscillating source, which would then result in a loss of resolution,
numerical instabilities, and an eventual breakdown of the code.
However, this turns out not to be the case;
all outgoing radiation {\em is} ``frozen'' around
$r\approx r_{\rm c}$, but the steep gradients which subsequently form 
in this region are efficiently and stably annihilated by the 
dissipation which is explicitly added to the difference scheme.

%\label{fig:dissplot2}
%\begin{figure}
%\epsfxsize=7.5cm
%\centerline{\epsffile{dissplot2.eps}}
%\caption[dissplot2]
%{
%
%}
%\label{fig:dissplot2}
%\end{figure}

In fact, there {\it is} a loss of resolution and
second order convergence for $r \sim r_{\rm c}$, but this does
not affect the stability or convergence of the solution for $r \ll r_{\rm c}$.  
Figure (\ref{fig:convtest}) shows a convergence test for the field
$\phi$ for $r < r_{\rm c}/2$ over roughly six crossing times.
%\label{fig:convtest}
%\begin{figure}
%\epsfxsize=7.5cm
%\centerline{\epsffile{4h2h_2hh.eps}}
%\caption[Convergence Test]
%{
%Convergence factor $||\phi_{4h} - \phi_{2h}||_2/||\phi_{2h}-\phi_{h}||_2$ 
%for the field $\phi$
%composed from the l2-norms of the solution at four different discretizations 
%(value of 4 indicates 2nd order convergence).
%}
%\label{fig:convtest}
%\end{figure}
Since we are solving equation (\ref{eq:boxphi}) in flat spacetime, it is very
simple to monitor energy conservation.  The spacetime 
admits a timelike Killing vector, $t^\nu$, so we have 
% $\nabla^\mu \left(t^\nu T_{\mu\nu}\right)= \nabla^\mu J_\mu=0$, yielding a
a conserved current, $J_\mu\equiv t^\nu T_{\mu\nu}$.  We monitor the flux of $J_\mu$ through
a surface constructed from two adjacent spacelike hypersurfaces
for $r\leq r_{\rm c}$ (with normals $n_\mu = (\pm 1,0,0,0)$),
and an ``endcap'' 
at $r=r_O$ (with normal
$n^\mu = (0,a^{-1},0,0))$.
%The normals to the Gaussian surface in MIB coordinates are 
%$n_\mu = (\alpha,0,0,0)$,
%$n_\mu = (-\alpha,0,0,0)$,
%and
%$n^\mu = (0,a^{-1},0,0)$
%for the top, bottom, and ``endcap'', respectively.
To obtain the the conserved energy at a time, $t_f$, 
the energy contained in the bubble,
\be
E_{\rm bubble} = 4\pi\hspace{-0.15in}\int\limits_{0}^{\ \ \ r_O} 
\hspace{-0.125in}r^2b^2 \left(
\frac{\Pi^2 + \Phi^2}{2 a^2} + V(\phi)\right) dr, 
\label{eq:ebub}
\ee
(where the integrand is evaluated at time $t_f$) 
is added to the total radiated energy, 
\be 
E_{\rm rad} = 4\pi\hspace{-0.05in}\int\limits_{0}^{\ t_f}
\hspace{-0.05in} r^2b^2 
\frac{\Pi \Phi}{a^2} dt
\label{eq:erad}
\ee
(where the integrand is evaluated at $r=r_O$). 
The sum, $E_{\rm total} = E_{\rm bubble} + E_{\rm rad}$, remains
conserved to within a few tenths of a 
percent\footnote{A few hundredths of a percent if measured relative
to the energy remaining after 
the initial radiative burst from the collapse.}
through a quarter million iterations (see Fig. (\ref{fig:econ_a})).
  
%\label{fig:econ_a}
%\begin{figure}
%\epsfxsize=7.5cm
%\centerline{\epsffile{econ.eps}}
%\caption[Energy Conservation]
%{
%Plot of energy contained in bubble (dashed lines), 
%energy radiated (dotted lines),
%and total energy (solid lines).  
%}
%\label{fig:econ_a}
%\end{figure}

Although monitoring energy conservation is a very important test,
it says little about whether there is reflection of the field off of the 
outer boundary, $r=r_{\rm max}$, or the region $r\approx r_{\rm c}$.
To check the efficacy of our technique for implementing  non-reflecting 
boundary conditions,  we compare the MIB results to those obtained 
with two other numerical schemes.
The first alternate method involves evolution of equation (\ref{eq:boxphi}) 
in ($\tilde{t}$,$\tilde{r}$) 
coordinates on a grid with $r_{\rm max}$ sufficiently large that 
radiation never reaches the outer boundary (large-grid solutions).  
For a given discretization scale,
results from this approach serve as
near-ideal reference solutions, since the solution
is guaranteed to be free of contamination from reflection
off the outer boundary.  The second method  
involves evolution on a grid with the same $r_{\rm max}$ adopted in 
the MIB calculation, but with discrete versions of massless
Sommerfeld (outgoing radiation) conditions applied at $r=r_{\rm max}$.
We refer to the results thus generated as OBC solutions, and since we
know that these solutions {\em do} have error resulting from reflections
from $r=r_{\rm max}$, they demonstrate what can go
wrong when a solution is contaminated by reflected 
radiation.
%\label{fig:lerr}
%\begin{figure}
%\epsfxsize=7.5cm
%\centerline{\epsffile{lerr_a.eps}}
%\caption[log absolute error]
%{
%Plot of $||\phi-\phi_{\rm ideal}||_2$ for MIB solution versus OBC
%solution.  
%%The error in the OBC solution increases dramatically at two crossing times. 
%}
%\label{fig:lerr}
%\end{figure}
Treating the large-grid solution as ideal, Fig. (\ref{fig:lerr}) 
compares typical $\log_{10}||\phi-\phi_{\rm ideal}||_2$ for 
the MIB and OBC solutions.
There is a steep increase in the OBC 
solution error (three orders of magnitude) 
around $t=125$, which is at roughly
two crossing times.  This implies that some radiation emitted from 
the initial collapse reached the outer boundary and reflected
back into the region $r<r_O$.  There is no such behavior
found in any MIB solutions.
% the difference between the MIB and
%large-grid solution grows roughly as $t^\alpha$ 
%for $\alpha$ around one (typical for second order
%codes with Kreiss-Oliger dissipation).
Lastly, for a more direct look at the field itself, we can see
$\phi(t,0)$ for large-grid (triangles), MIB (solid curves), and OBC 
(dashed curves) solutions 
in Fig. (\ref{fig:crossingtimes}).
%\label{fig:crossingtimes}
%\begin{figure}
%\epsfxsize=7.5cm
%\centerline{\epsffile{crossingtimes.eps}}
%\caption[phi(0,t) after many crossing times]
%{
%$\phi(t,r\!\!=\!\!0)$ for the large-grid solution (triangles),
%MIB solution (solid curves), and OBC solution (dashed curves).
%The solutions all agree before $2 t_{\rm crossing}$, but the
%OBC solution begins to drift away from the ideal solution
%after $2 t_{\rm crossing}$.  All pictures span the same area, 
%$\Delta \phi = 0.075$ by $\Delta t = 0.5$. 
%}
%\label{fig:crossingtimes}
%\end{figure}
Initially, both the MIB and OBC solutions agree with the large-grid 
solution extremely well. However, after two crossing times the OBC solution 
starts to substantially diverge from the ideal solution, while the 
MIB results remain in very good agreement with the ideal calculations.

In summary, the MIB solution conserves energy,  
converges quadratically in the mesh spacing (as expected), and produces 
results which are equivalent---at the level of truncation error---to
large-grid reference solutions.  At the same time, the MIB approach
is considerably more computationally efficient than dynamical- or 
large-grid techniques.

%**********************************************************************
%\begin{references}

%--------------------------- TABLE ---------------------------
\newpage

\begin{table}[ht]
%\label{tab:fdop}
\begin{tabular}{lcr}
\colrule
Operator & Definition & Expansion \\
\colrule
$\Delta^f_r f^n_i$ & $\left( -3f^n_i + 4f^n_{i+1}-f^n_{i+2}\right) /
2\Delta r$ &
$\partial_r f \big\vert^n_i + O\left(\Delta r^2\right)$ \\
$\Delta^b_r f^n_i$ & $\left( 3f^n_i - 4f^n_{i-1} + f^n_{i-2}\right)
/2\Delta r$ &
$\partial_r f \big\vert^n_i + O\left(\Delta r^2\right)$ \\
$\Delta_r f^n_i$ & $\left( f^n_{i+1}-f^n_{i-1}\right) /2\Delta r$ & 
$\partial_r f \big\vert^n_i + O\left(\Delta r^2\right)$ \\
$\Delta_t f^n_i$ & $\left( f^{n+1}_i-f^n_i\right)/\Delta t$ &
$\partial_t f \big\vert^{n+\frac{1}{2}}_i + O\left( \Delta t^2\right)$ \\
$\Delta^d_t f^n_i$ & $\left( f^{n+1}_i-f^n_i\right)/\Delta t + $ &
$\partial_t f \big\vert^{n+\frac{1}{2}}_i + O\left( \Delta t^2\right)$ \\
 & $\epsilon_{dis}[ 6f^n_i + f^n_{i-2}+f^n_{i+2} - $ & \\
 & $4\left( f^n_{i-1}+f^n_{i+1}\right)] /16\Delta t$ & \\
$\mu_t f^n_i$ & $\left( f^{n+1}_i + f^n_i\right) / 2$ &
$f\big\vert^{n+\frac{1}{2}}_i + O\left( \Delta t^2\right)$ \\
\colrule
\end{tabular}
\vspace{1cm}
\caption{\label{tab:fdop}Two-Level Finite Difference Operators.  Here 
we have adopted a standard finite-difference notation: $f^n_i \equiv 
f((n-1)\Delta t,(i-1)\Delta r)$}
\end{table}

% ------------------------ FIGURES ----------------------------
\newpage

\newpage

\begin{figure}
\centerline{ 
	\hbox{\epsfxsize =12.5cm\epsffile{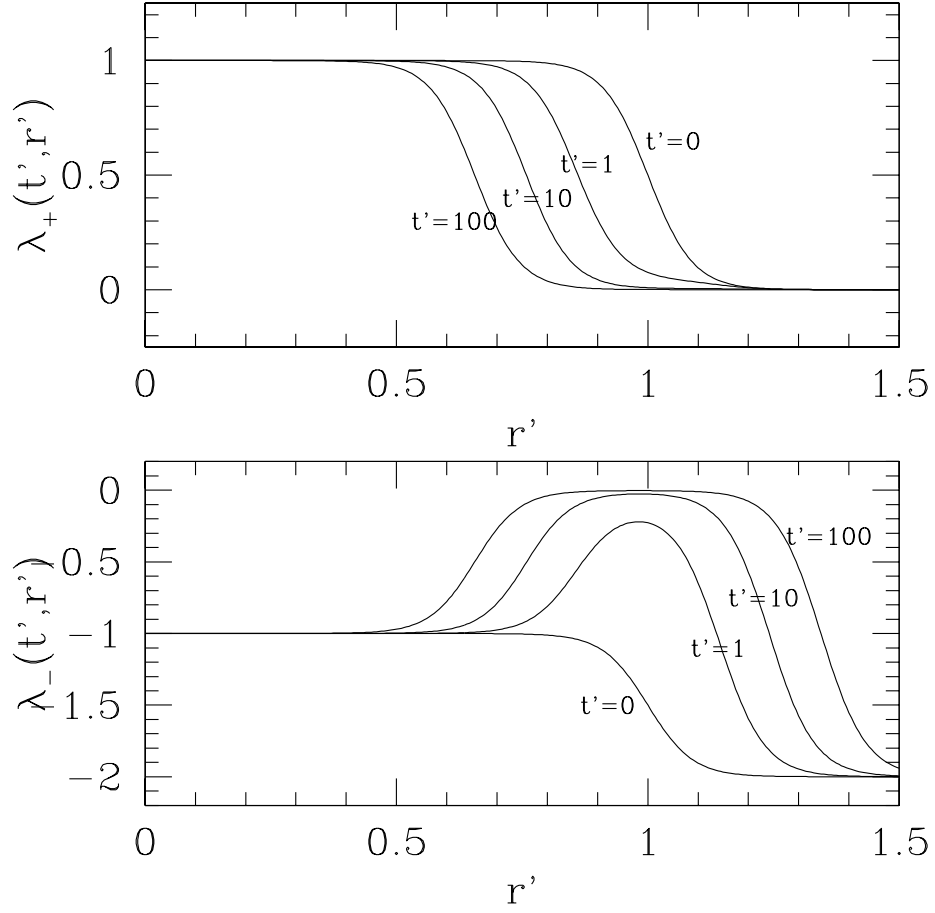}}
	}
\vspace{1cm}
\caption[Characteristics for radial MIB system]
{\small \label{fig:1Dchars} 
Plot of the characteristic velocities as a function of the 
MIB coordinates, $r'$ and 
$t'$ in units where $r_c$ is set to unity. $\lambda_+$ and $\lambda_-$ are the outgoing and 
ingoing characteristic speeds
respectively.  $\delta$ is taken to be $\delta\rightarrow \delta/r_c \approx 0.0893$ 
(corresponding to the system used in this article).
Characteristic velocities are plotted for times $t'=0$, $1$, $10$, and $100$
($t'=100$ is larger than the lifetime of the longest lived solution studied in this work).
As $r\to1$, we have $\lambda_\pm(t',r') = O(1/t')$.
}
%\label{fig:1Dchars}
\end{figure}

\newpage

\begin{figure}
\centerline{
        \hbox{\epsfxsize =12.5cm\epsffile{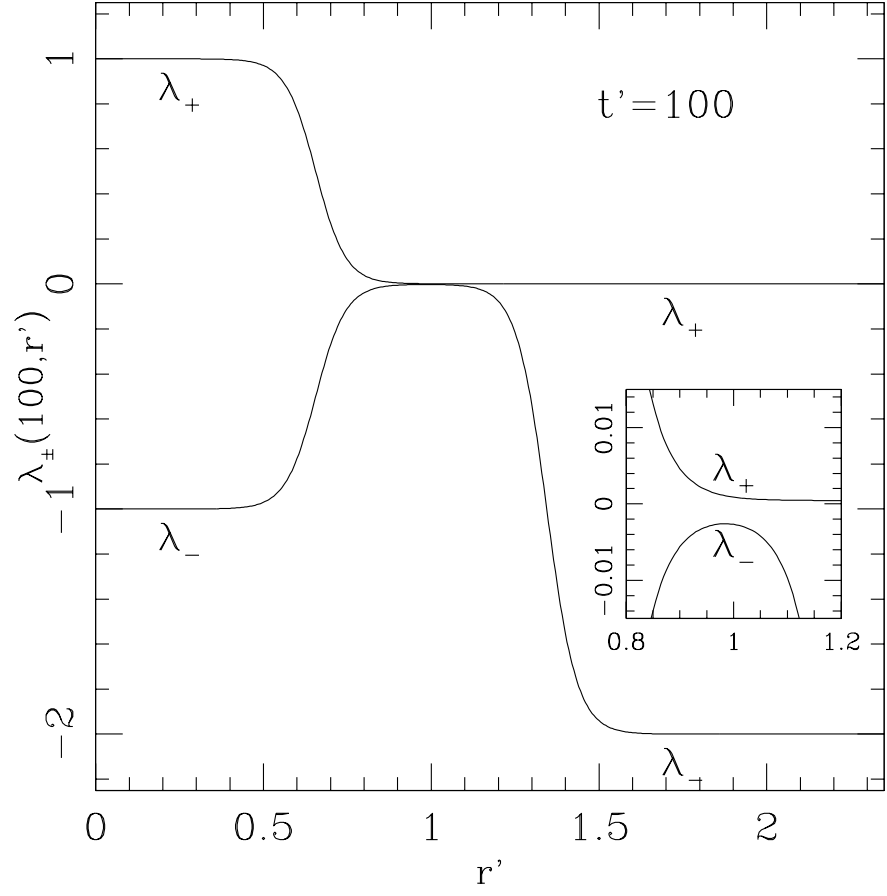}}
        }
\vspace{1cm}
\caption[Coincidence of ingoing/outgoing $r$:$t$ characteristics]
{\small \label{fig:1Dchars_coin}
Plot of characteristic speeds, $\lambda_\pm(r',100)$,
where $r'$ and $t'$ are radial MIB coordinates in units where $r_c$ is set to
unity.
}
%\label{fig:1Dchars_coin}
\end{figure}

\newpage

\begin{figure}
\epsfxsize=12.5cm
\centerline{\epsffile{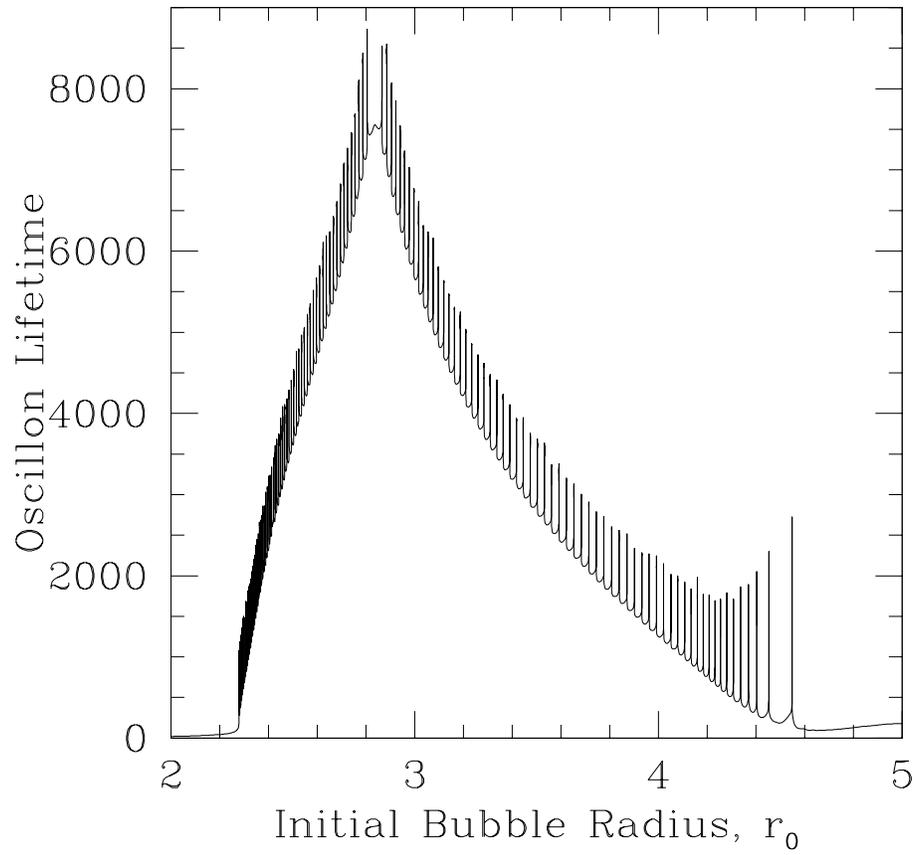}}
\vspace{1cm}
\caption{ 
Plot of oscillon lifetime versus initial bubble radius
for $2.0 \le r_0 \le 5.0$.
Each of the 125 resonances is resolved to one part in $10^{14}$.
%With the high resolution parameter space survey the 125 resonances
%become apparent.  The survey was conducted at (relatively)
%coarse resolution to reveal the location of the resonances.
%Once a resonance was found, it was resolved efficiently
%first by using a three-point routine that maximizes
%the lifetime as a function of initial bubble radius, and then by
%bisecting (to one part in $10^{14}$)
%on the bifurcate modulation behavior discussed below
%(and seen in figure \ref{fig:phenv_energy}).
}
\label{fig:lifetime1}
\end{figure} 

\newpage

\begin{figure}
\epsfxsize=12.5cm
\centerline{\epsffile{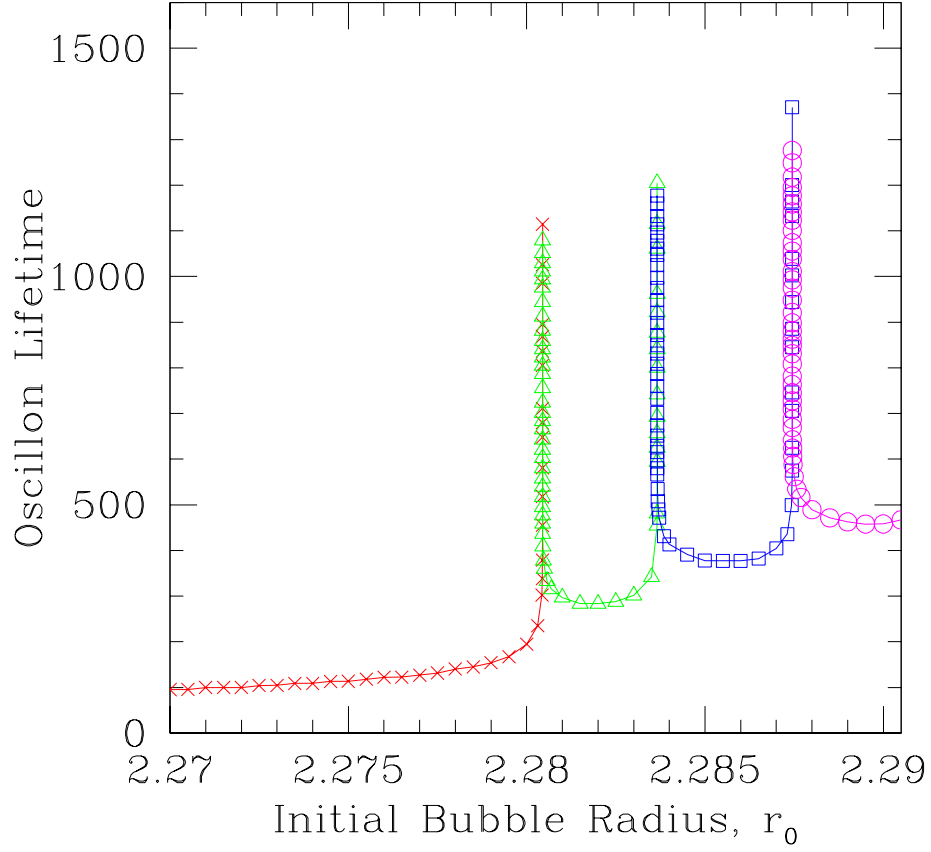}}
\vspace{1cm}
\caption[Oscillon lifetime versus initial bubble radius]
{
Plot of oscillon lifetime versus initial bubble radius
for  $2.27\le r_{\rm initial}\le 2.29$.
The three resonances shown occur at
$r^*_0\approx 2.2805$,
$r^*_1\approx 2.2838$,
and
$r^*_2\approx 2.2876$.
Each resonance separates the parameter space into regions
with $n$ and $n+1$ modulations; the x's correspond
to oscillons with no modulations, the triangles to oscillons
with one modulation, the squares to two modulations, and the
circles to three modulations.
}
\label{fig:life_1449.r0=2.27_2.29}
\end{figure}

\newpage

\begin{figure}
\epsfxsize=12.5cm
\centerline{\epsffile{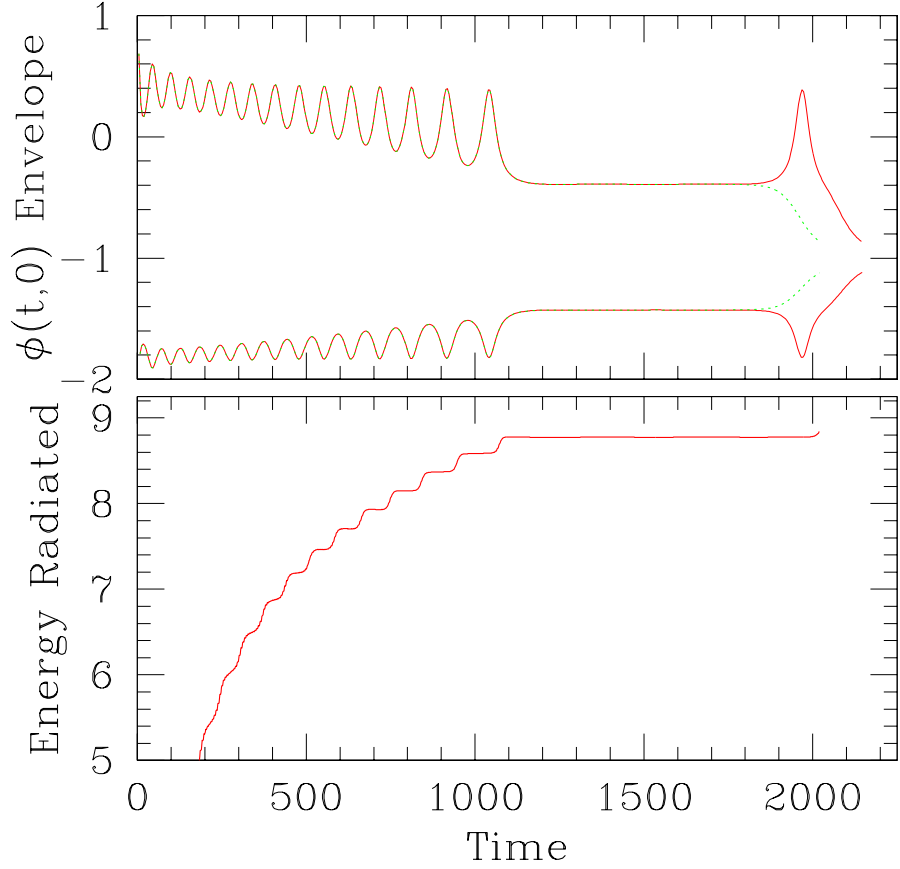}}
\vspace{1cm}
\caption[Field envelope barely above/below resonance and
power radiated versus time. ]
%%%%%%%%%%%%%%%%%%%%%%%%%%%%%%%%%%%%%%%%%%%%%%%%%%%%%%%%%%%%%%%%
%%%%%%%%%%%%%%%%%%%%%%%%%%%%%%%%%%%%%%%%%%%%%%%%%%%%%%%%%%%%%%%%
{\small \label{fig:phenv_energy}
Top plot shows the envelope of $\phi(t,0)$ for $r^\star_0\!\pm\!\Delta r_0$
displaying bifurcate behavior around the
$r^\star_0 \approx 2.335$ resonance ($\Delta r_0 \sim 10^{-14}$);
the solid curve is the envelope barely above resonance 
(15 modulations) while
the dotted line is the envelope barely below resonance
(14 modulations).
Bottom plot shows the energy radiated as a function of time 
through the surface containing the oscillon as defined in Appendix B.
The increases in the
energy radiated are synchronized with the modulation in the field.
}
%\label{fig:phenv_energy}
\end{figure}

\newpage

\begin{figure}
\epsfxsize=12.5cm
\centerline{\epsffile{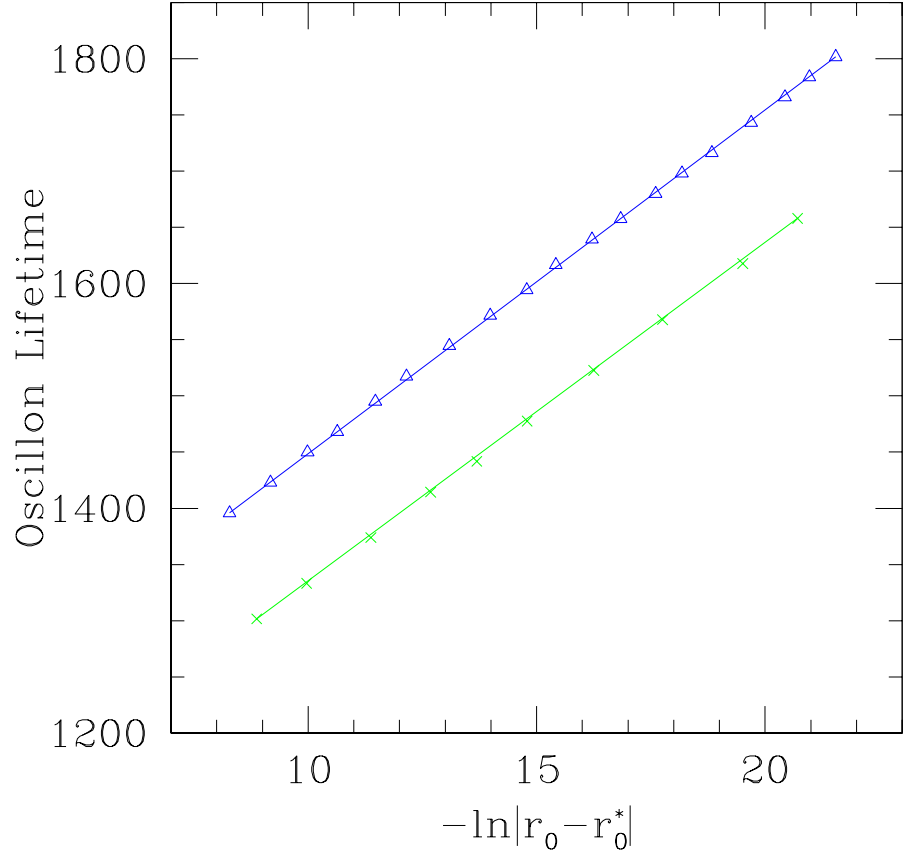}}
\vspace{1cm}
\caption[Time scaling, $T$ versus $-\ln|r_0 - r_0^\star|$ about a resonance]
{\small \label{fig:tscaling}
Plot of time scaling, $T$ versus $\ln|r_0 - r_0^\star|$ for the
 $r_0\approx 2.335$ resonance.
The top line (triangles) displays the scaling behavior for supercritical
evolutions, $r_0>r_0^\star$,
while the bottom line (x's) shows the scaling for subcritical
calculations, $r_0<r_0^\star$.
The exponents (measured by the slopes of the lines) are both approximately 
equal to $\gamma = 30$.
The offset in the two curves 
represents the time spent by supercritical oscillons 
in executing the final modulation shown in Fig.~6.
}
%\label{fig:tscaling}
\end{figure}

\newpage

\begin{figure}
\epsfxsize=12.5cm
%\centerline{\epsffile{gamma_life_vs_r.eps}}
\centerline{\epsffile{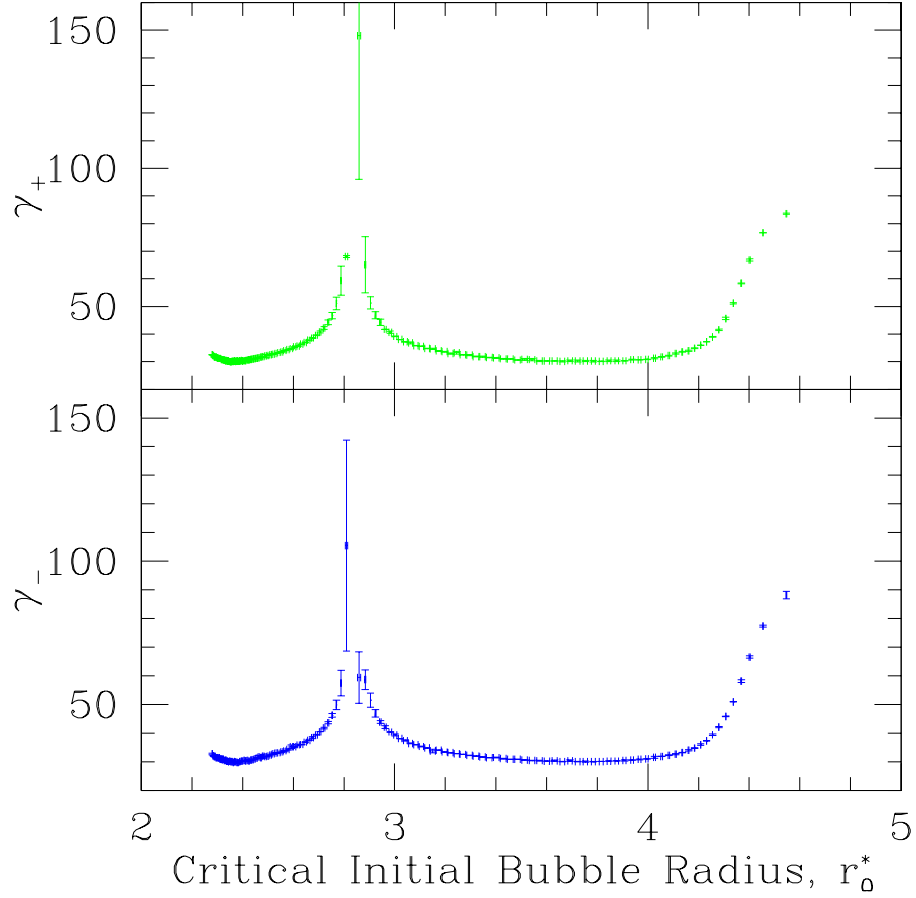}}
\vspace{1cm}
\caption[Critical exponents for each resonance]
{\small \label{fig:exponents}
Plot of critical exponents for each resonance.
There are two values of $\gamma$ for each resonance. The top plot displays
the $\gamma_+$ vs. $r_0^\star$ while the lower plot displays
$\gamma_-$ vs. $r_0^\star$.
The uncertainties are estimated from running the {\it entire} parameter
space surveys at two resolutions, $N\equiv N_r=1449$ and $N' \equiv N_r'=1025$ and estimating the error,
$\Delta\gamma = |\gamma_{N} -\gamma_{N'}|$.  
}
%\label{fig:exponents}
\end{figure}

\newpage

\begin{figure}
\epsfxsize=12.5cm
\centerline{\epsffile{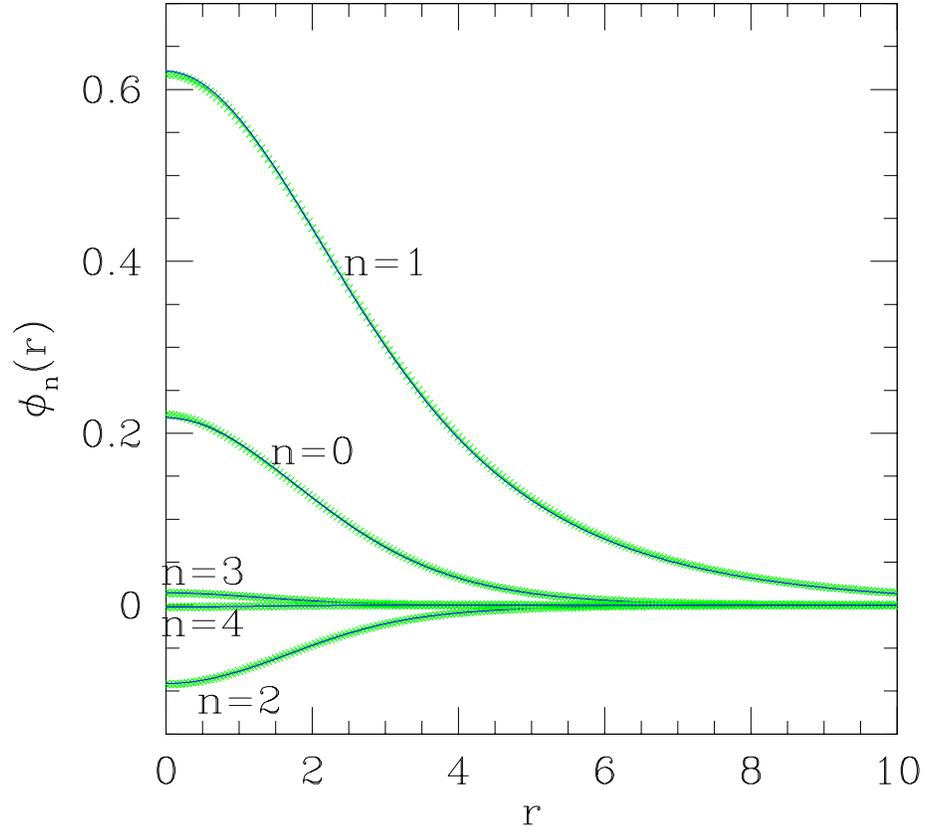}}
\vspace{1cm}
\caption[Mode decomposition of critical oscillon]
{\small \label{fig:modstruct}
Critical solution  $\phi_n(r)$ (for $n=0,1,2,3,4$) obtained
from the Fourier-decomposed PDE data (x's)
overlayed with $\phi_n(r)$ obtained by
shooting equations \ref{modestruct0} and \ref{modestructn}
(solid curves).
The Fourier-decomposed PDE data overlays the
shooting solution everywhere.
%The dynamic data was Fourier decomposed by forming
%a time series for each spatial gridpoint, $r_i$, while
%the oscillon entered its non-radiative (no modulation) phase.
%FFT's were then constructed for each time series and the amplitude of
%each mode  (the green x's) was measured.
}
%\label{fig:modstruct}
\end{figure}

\newpage

\begin{figure}
\epsfxsize=12.5cm
\centerline{\epsffile{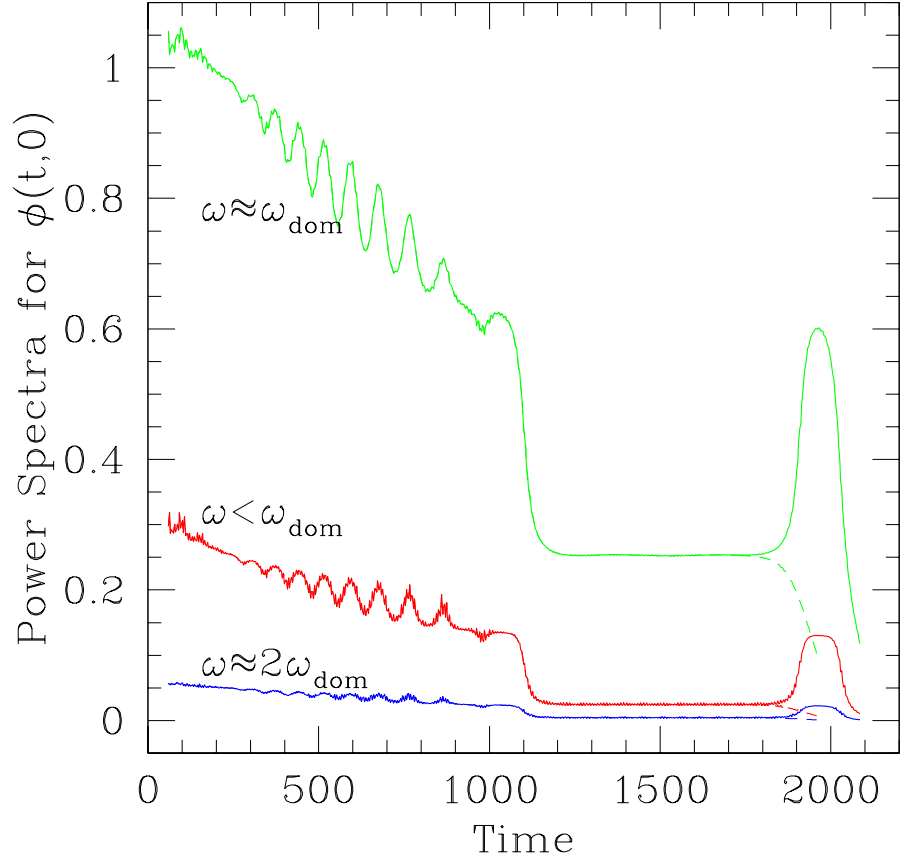}}
\vspace{1cm}
\caption[Power spectra of core amplitude versus time at resonance]
{
Power spectra of the core amplitude, $\phi(t,0)$,
for the oscillons barely above and below the
$r_0\approx2.335$ resonance.
The power measured in each frequency regime slowly diminishes
as the oscillon radiates away much of its energy until
approximately $t=1100$ where the oscillon enters a non-radiative
state and all the components of the power spectrum become
constant.
}
\label{fig:modrad}
\end{figure}

\newpage

\begin{figure}
\epsfxsize=12.5cm
\centerline{\epsffile{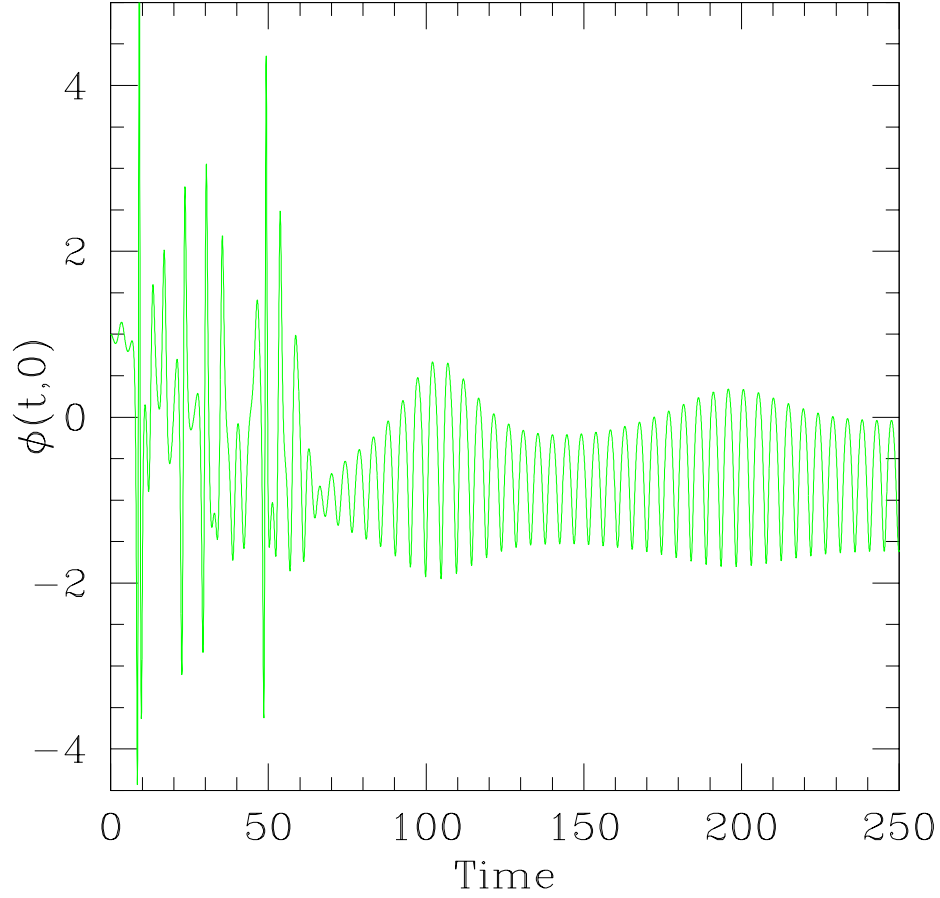}}
\vspace{1cm}
\caption[Chaotic behavior of $\phi(t,0)$ for ``bouncing'' bubble]
{
Plot of $\phi(t,0)$ for $r_0=7.25$ displaying extremely nonlinear and
unpredictable behavior during the ``bouncing'' phase (for $t< 60$),
after which the field settles into a typical oscillon evolution.
Once in the oscillon regime, the period
is approximately $T \approx 4.6$, and the first two modulations
of the field can be seen (envelope
maxima at $t\approx 105$ and $t\approx 200$).
}
\label{fig:corebounce} %EPH2: had no label?
\end{figure}

\newpage

\begin{figure}
\epsfxsize=12.5cm
\centerline{\epsffile{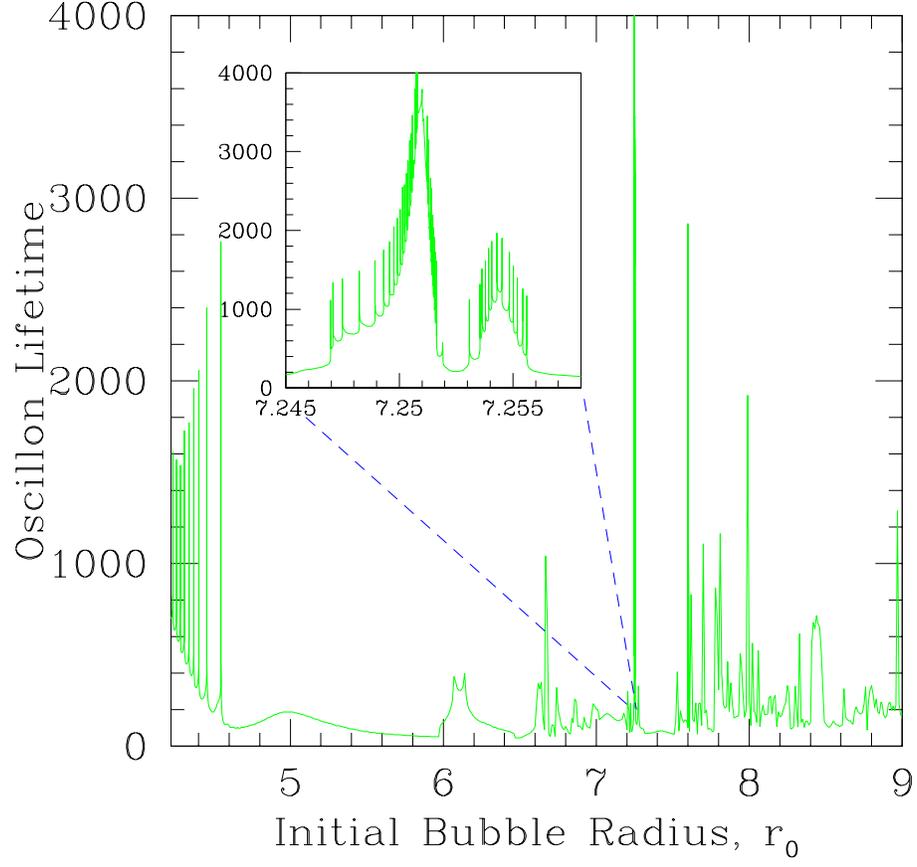}}
\vspace{1cm}
\caption[Oscillon lifetime versus initial bubble radius in bounce regime]
{
Plot of oscillon lifetime versus initial radius of bubble
for $4.22\le r_0\le 9$.
Although there seem to be no oscillons in the range 
$4.6\lesssim r_0\lesssim 6$, it is clear that
oscillons and resonances {\it do} exist for higher initial bubble radii,
$r_0 \gtrsim 6.5$.
}
\label{fig:lifefine}
\end{figure}

\newpage

%%%%%%%%%%%%%%%%%%%%%%%%%%%%%%%%%%%%%%%%%%%%%%%%%%%%%%%%%%%%%%%%
%%%%%%%%%%%%%%%%%%%%%%%%%%%%%%%%%%%%%%%%%%%%%%%%%%%%%%%%%%%%%%%%
\begin{figure}
\epsfxsize=12.5cm
\centerline{\epsffile{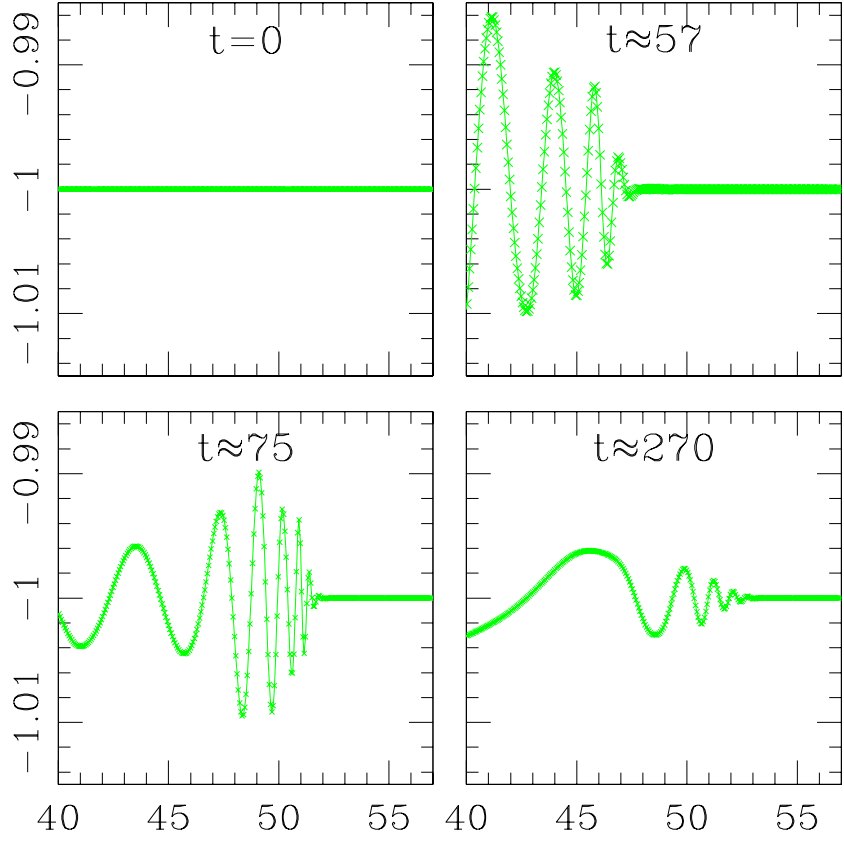}}
\vspace{1cm}
\caption[Outgoing radiation in freeze-out region]
{\small \label{fig:dissplot2}
Fundamental field $\phi(r)$ in the freeze-out region
at $t=0,\ 57,\ 75,\  {\rm and}\ 270$.
The characteristic speeds of the radiation $\to 0$ as $r \to r_c$
(here, $r_c = 56$)
%(as seen in figure \ref{fig:1Dchars_coin}),
and the wavelength of the radiation
is blue-shifted to the lattice Nyquist limit, $2\Delta r$.
The Kreiss-Oliger dissipation explicitly added to the finite difference
equations
subsequently ``quenches'' the field.
%figure \ref{ampfacs} shows the amplification factor $|\rho|^2$
%as a function of wavenumber is significantly less than one as the wavenumber
%approaches the Nyquist limit (the plot of $|\rho(\xi)|^2$
%is actually for the advection equation, but is qualitatively similar).
}
%\label{fig:dissplot2}
\end{figure}

\newpage

\begin{figure}
\epsfxsize=12.5cm
\centerline{\epsffile{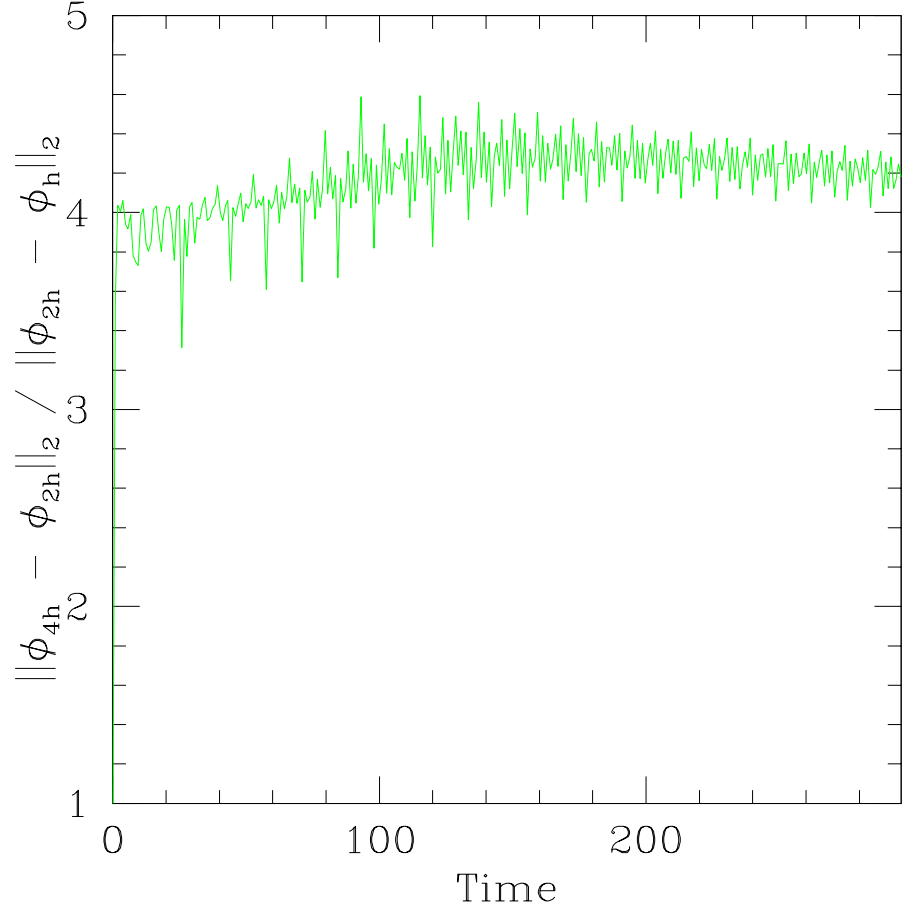}}
\vspace{1cm}
\caption[Convergence test of fundamental field, $\phi$]
{\small \label{fig:convtest}
Convergence factor, $C_f=||\phi_{4h} - \phi_{2h}||_2/||\phi_{2h}-\phi_{h}||_2$,
for the field $\phi$
composed from the solution at three different discretizations
%(value of 4 indicates 2nd order convergence, as shown in
%section \ref{sec:convergence}). 
(value of 4 indicates 2nd order convergence). The $\ell_2$ norm $||\cdots||_2$ 
is defined by $||v||_2 = \left( N^{-1} \sum_{i=1}^{N} v_i \right)^{1/2}$
}
%\label{fig:convtest}
\end{figure}

\newpage

\begin{figure}
\epsfxsize=12.5cm
\centerline{\epsffile{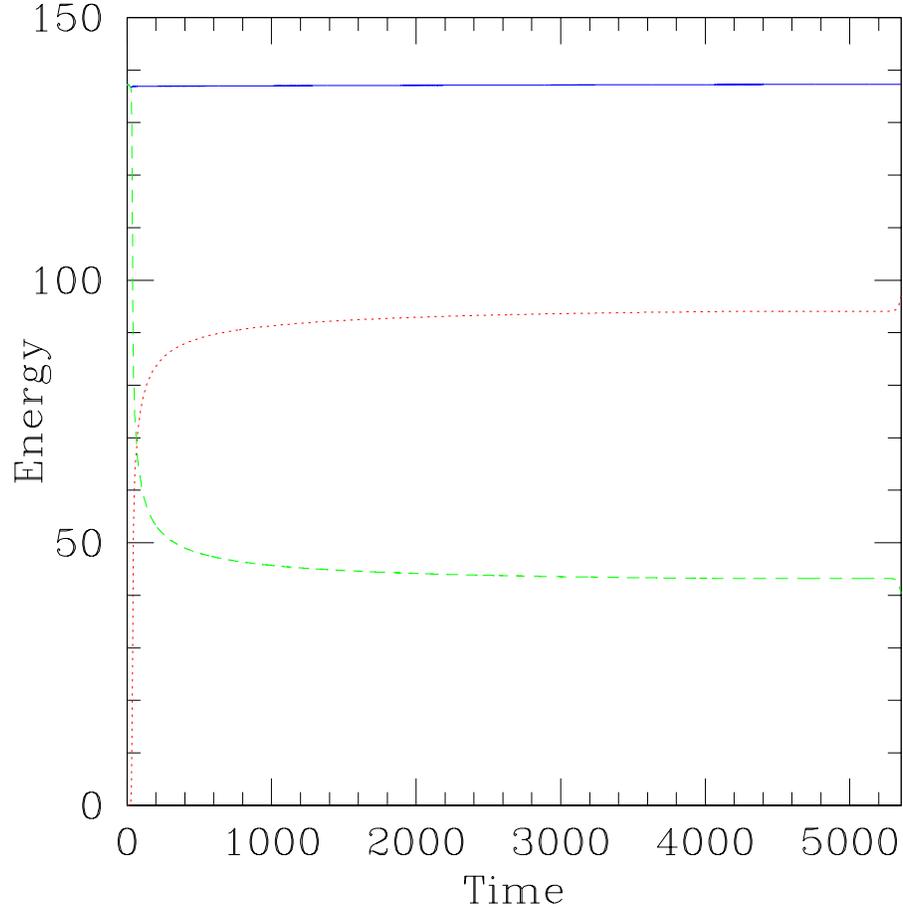}}
\vspace{1cm}
\caption[Energy conservation test of 1D bubble code]
{\small \label{fig:econ_a}
Plot of energy contained in oscillon (dashed line),
energy radiated (dotted line),
and total energy (solid line).
The total energy of the system is a constant of motion and is numerically
conserved to within a few tenths of a percent.
The energy contained within the bubble drops rapidly during the
initial radiative phase and plateaus around $E\approx 43m/\lambda$
during the quasi-stable ``oscillon'' phase.
}
%\label{fig:econ_a}
\end{figure}

\newpage

\begin{figure}
\epsfxsize=12.5cm
\centerline{\epsffile{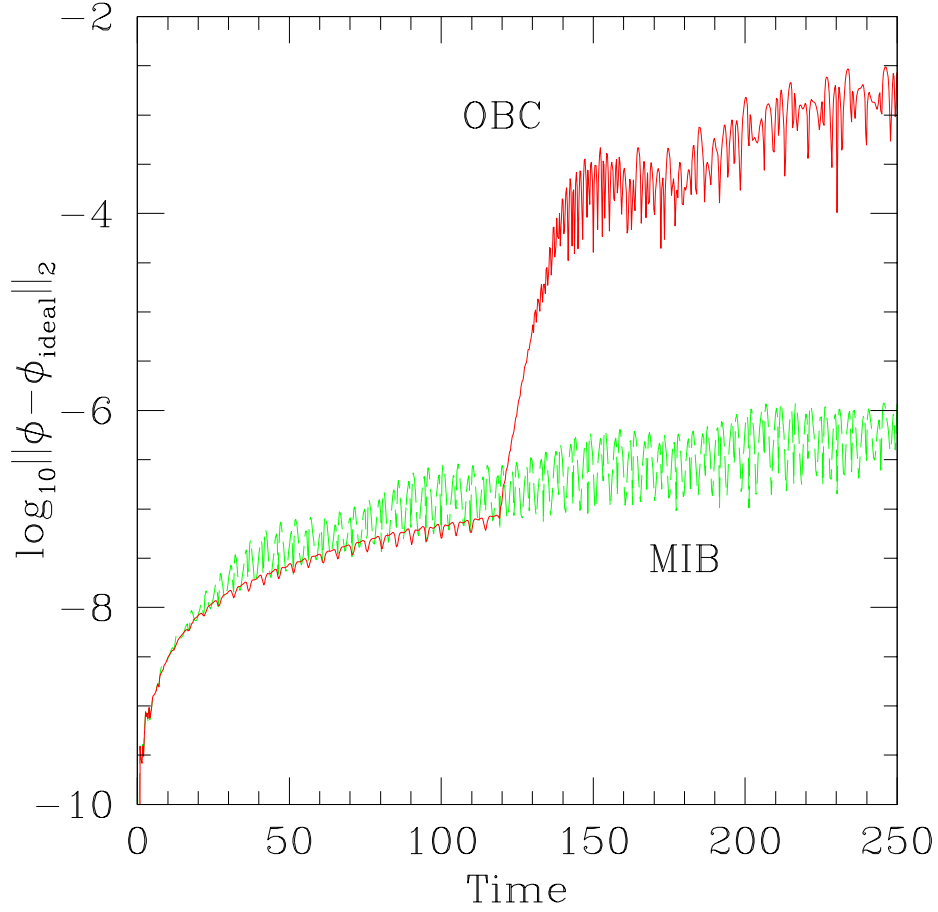}}
\vspace{1cm}
\caption[Contamination test of 1D bubble code]
{ \small \label{fig:lerr}
Plot comparing the OBC (solid line) and MIB (dashed line)
solutions to an ``ideal'' solution.
The OBC solution is obtained using a massless outgoing
boundary condition, the MIB solution is obtained by solving the
system in spherical MIB coordinates, and the ideal solution is
obtained by evolving the solution in standard ($r$,$t$) coordinates on
a grid large enough to ensure no reflection off the outer boundary.
The error estimates are obtained from the $\ell_2$-norm of the difference
between the trial solutions (OBC or MIB) and the ideal solution,
$||\phi-\phi_{\rm ideal}||_2$.
Contamination of the OBC solution is observed at two crossing times, $t\approx 120$,
where the error estimate increases by over three orders of magnitude.
The MIB solution error grows slowly and steadily as expected when solving
a continuum equation with two {\it different} finite difference
techniques.
}
%\label{fig:lerr}
\end{figure}

\newpage

\begin{figure}
\epsfxsize=12.5cm
\centerline{\epsffile{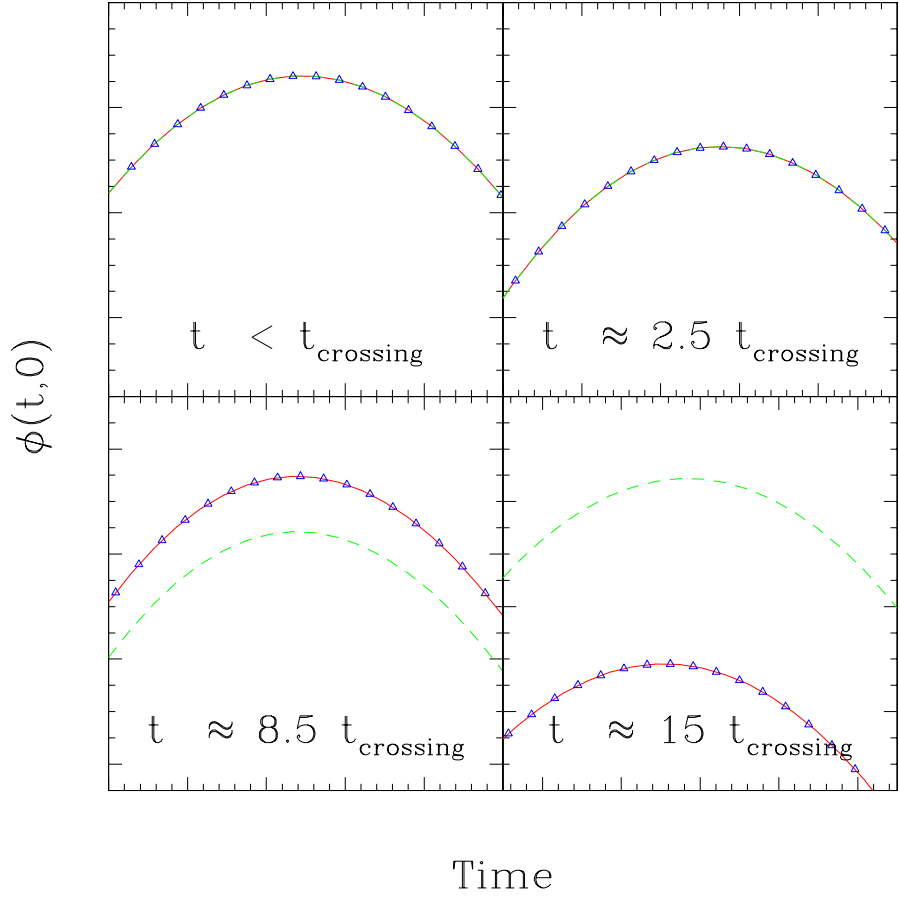}}
\vspace{1cm}
\caption[Contamination test for 1D bubble code, $\phi(t,0)$
 after many crossing times]
{
$\phi(t,0)$ versus time
for the large-grid solution (triangles),
MIB solution (solid curves), and OBC solution (dashed curves).
The solutions all agree before $2 t_{\rm crossing}$, but the
OBC solution begins to drift away from the ideal solution
after $2 t_{\rm crossing}$.
The error in the OBC solution is due to radiation that is
reflected off of the outer boundary (hence needing two crossing
times to return to $r=0$ to contaminate the oscillon).
All pictures span the same area,
$\Delta \phi = 0.075$ by $\Delta t = 0.5$.
}
\label{fig:crossingtimes}
\end{figure}

%\cleardoublepage

\end{document}